\documentclass[11pt]{article}
\usepackage[dvipdfmx]{color}
\usepackage{graphicx}
\usepackage{xcolor}
  \usepackage{amsthm,amsfonts}
  \usepackage{amsmath}
\usepackage{mathabx}
\usepackage{slashed}
\usepackage{ulem}
\usepackage{cite}

\newcommand{\bea}   {\begin{eqnarray}}
\newcommand{\eea}   {\end{eqnarray}}

\def\nn{\nonumber}

\def\pl#1{\overrightarrow{\partial}_{#1}}
\def\pr#1{\overleftarrow{\partial}_{#1}}
\def\PB#1{ \{ #1 \}_{\textrm{\tiny P}}}
\def\DB#1{ \{ #1 \}_{\textrm{\tiny D}}}
\begin{document}
\renewcommand{\thefootnote}{\fnsymbol{footnote}}

\thispagestyle{empty}

\title{${\mathbb Z}_2\times {\mathbb Z}_2$-graded mechanics: the quantization}
\author{N. Aizawa\thanks{{E-mail: {\it aizawa@p.s.osakafu-u.ac.jp}}}, \quad Z. Kuznetsova\thanks{{E-mail: {\it zhanna.kuznetsova@ufabc.edu.br}}}\quad and\quad
F.
Toppan\thanks{{E-mail: {\it toppan@cbpf.br}}}
\\
\\
}
\maketitle

\centerline{$^{\ast}$ {\it Department of Physical Sciences, Graduate School of Science,}}{\centerline{\it  Osaka Prefecture University, Nakamozu Campus,}}{\centerline{\it Sakai, Osaka 599-8531 Japan.}}
\centerline{$^{\dag}$ {\it UFABC, Av. dos Estados 5001, Bangu,}}\centerline{\it { cep 09210-580, Santo Andr\'e (SP), Brazil.}}
{\centerline{$^{\ddag}$ 
{\it CBPF, Rua Dr. Xavier Sigaud 150, Urca,}}\centerline{\it{
cep 22290-180, Rio de Janeiro (RJ), Brazil.}}
~\\
\maketitle
\begin{abstract}
In a previous paper we introduced the notion of ${\mathbb Z}_2\times{\mathbb Z}_2$-graded classical mechanics and presented a general framework to construct, in the Lagrangian setting, the worldline sigma models
invariant under a ${\mathbb Z}_2\times{\mathbb Z}_2$-graded superalgebra. In this work we discuss at first the classical Hamiltonian formulation for a class of these models and later present their canonical quantization. \par
As the simplest application of the construction we recover the ${\mathbb Z}_2\times{\mathbb Z}_2$-graded quantum Hamiltonian introduced by Bruce and Duplij. We prove that this is just the first example of a large class of ${\mathbb Z}_2\times{\mathbb Z}_2$-graded quantum models. We derive in particular interacting multiparticle  quantum Hamiltonians given by Hermitian, matrix, differential operators. The interacting terms appear as non-diagonal entries in the matrices. \par The construction of the Noether charges, both classical and quantum, is presented. A comprehensive discussion of the different ${\mathbb Z}_2\times{\mathbb Z}_2$-graded symmetries possessed by the quantum Hamiltonians is given.

\end{abstract}
\vfill
\rightline{CBPF-NF-003/20}
\newpage

\section{Introduction}

${\mathbb Z}_2\times{\mathbb Z}_2$-graded Lie superalgebras \cite{{rw1},{rw2}} are generalizations of ordinary Lie superalgebras. For many years they have been a research topics in mathematics, in physics and in application to parastatistics. Despite this activity, a systematic investigation of models with ${\mathbb Z}_2\times{\mathbb Z}_2$-graded symmetry  was not available. In a previous paper \cite{AKT1} we constructed classical theories invariant under a ${\mathbb Z}_2\times{\mathbb Z}_2$-graded symmetry. They are formulated for a set of ${\mathbb Z}_2\times{\mathbb Z}_2$-graded fields: ordinary ($00$-graded) bosons, exotic ($11$-graded) bosons
and two classes of fermions ($01$- and $10$-graded), with fermions belonging to different classes which mutually commute instead of anticommuting.  \par
The canonical quantization of ${\mathbb Z}_2\times{\mathbb Z}_2$-graded models is not straightforward. One should be careful in the introduction of left\slash right derivations,  in defining Poisson brackets and related structures  of the Hamiltonian framework, since they have to be in accordance with the ${\mathbb Z}_2\times{\mathbb Z}_2$-grading.\par
In this paper we establish the canonical quantization procedure, extending the approach of \cite{CHT} for supermechanics, and apply it to derive a class of  ${\mathbb Z}_2\times{\mathbb Z}_2$-graded invariant quantum Hamiltonians. In particular we quantize the classical models
for a single $(1,2,1)_{[00]}$ multiplet of fields (this multiplet produces a propagating ordinary boson, an auxiliary exotic boson and a fermion in each one of the two classes) and for $n$ interacting $(1,2,1)_{[00]}$ multiplets. 
The quantization of the single ($n=1$) multiplet reproduces the ${\mathbb Z}_2\times{\mathbb Z}_2$-graded quantum mechanics introduced in\cite{BruDup}. New quantum models are obtained for $n>1$. They have interesting features that cannot be observed for $n=1$. The interacting terms among the multiplets appear as non-diagonal entries in their hermitian matrices. The $8\times 8$ differential matrix Hamiltonian for $n=2$ and the $16\times 16$ differential matrix Hamiltonian for $n=3$ are respectively presented in formulas (\ref{ham2p}) and (\ref{ham3pint}-\ref{hoff3p}). They are given in terms of the unconstrained functions $f(x,y)$ for $n=2$ and $f(x,y,z)$ for $n=3$.\par
${\mathbb Z}_2\times{\mathbb Z}_2$-graded (super)algebras were introduced by Rittenberg and Wyler in \cite{{rw1},{rw2}}, while earlier related structures
were investigated in\cite{Ree}. In the Rittenberg-Wyler works, which were inspired by the construction of ordinary superalgebras, some possible physical applications to elementary particle physics were suggested. Since then these new graded structures attracted the attention of mathematicians \cite{sch} with a steady flow of papers devoted to their classifications
\cite{{Sil},{SZZ}}, representations \cite{{Sil2},{ChSiVO},{SigSil}}, generalizations \cite{CART}. On the physical side ${\mathbb Z}_2\times{\mathbb Z}_2$-graded structures received some attention, see the works \cite{{lr},{jyw}, {zhe}, {Toro1}, {Toro2}} and, in connection with de Sitter supergravity, \cite{{vas},{tol2}}.\par
${\mathbb Z}_2\times{\mathbb Z}_2$-graded superalgebras naturally lead to the broad field of parastatistics (for the mathematical aspects of the connection with parastatistics, see \cite{{tol},{StVdJ}}). It is therefore quite natural to expect that they could play a role in low-dimensional (where anyons can enter the game) and/or non-relativistic physics.  This recognition is responsible for the recent surge of interest to physical applications of ${\mathbb Z}_2\times{\mathbb Z}_2$-graded superalgebras, with several works investigating this problem from different sides. In \cite{{aktt1},{aktt2}} it was shown, quite unexpectedly, that ${\mathbb Z}_2\times{\mathbb Z}_2$-graded superalgebras are symmetries of the well-known L\'evy-Leblond equations. 
The  ${\mathbb Z}_2\times {\mathbb Z}_2$-graded} analogues of supersymmetric and superconformal quantum mechanics were introduced in \cite{Bru,BruDup,NaAmaDoi,NaAmaDoi2}. Graded structures with commuting fermions appear in dual double field theory and mixed symmetry tensors \cite{{BHPR},{CKRS},{BruIbar}}. In the meanwhile, mathematical
properties continue to be investigated \cite{{AiIsSe},{BruDup2},{ISVdJ},{Me}}.\par
This state of the art motivated us to launch a systematic investigation of the features of ${\mathbb Z}_2\times {\mathbb Z}_2$-graded mechanics. In the first paper in this direction \cite{AKT1} we presented the general framework for
the construction of ${\mathbb Z}_2\times {\mathbb Z}_2$-graded invariant, classical, worldline sigma models in the Lagrangian setting. We mimicked the construction of supermechanics, that is the classical supersymmetric mechanics, extending the tools (the  $D$-module representations of supermultiplets and their application to the derivation of the invariant actions) developed in \cite{{PaTo},{KRT},{KT}}.\par
In the present paper devoted to the canonical quantization, besides the already mentioned results, we show in particular, by analyzing the
Noether charges, that both the one-dimensional ${\mathbb Z}_2\times{\mathbb Z}_2$-graded supertranslation algebra and the Beckers-Debergh algebra introduced in \cite{BecDeb} are obtained by taking into account different matrix representations for the (anti)commutators induced by the Dirac brackets.\par
The scheme of the paper is the following. In Section {\bf 2}, after recalling the Lagrangian formulation of the ${\mathbb Z}_2\times{\mathbb Z}_2$-graded classical invariant models under consideration, we derive their classical Noether charges. In Section {\bf 3} we present the Hamiltonian formulation of these classical models; the ``constant kinetic basis", which paves the way for the canonical quantization, is introduced. The canonical quantization and the derivation of the conserved
quantum Noether charges is presented in Section {\bf 4}. The different graded symmetries of the simplest quantum Hamiltonian are discussed in Section {\bf 5}. In Section {\bf 6} we present  invariant, ${\mathbb Z}_2\times {\mathbb Z}_2$-graded, interacting multiparticle quantum Hamiltonians. In the Conclusions we discuss the relevance of the results obtained in the paper and point out various directions of future works. For completeness, the relevant features of the ${\mathbb Z}_2\times{\mathbb Z}_2$-graded superalgebras and of their graded representations are recalled in the Appendix.

\section{The ${\mathbb Z}_2\times{\mathbb Z}_2$-graded classical Lagrangian mechanics}

We revisit at first the simplest cases of ${\mathbb Z}_2\times{\mathbb Z}_2$-graded classical mechanics in the Lagrangian formulation.  Later, at the end of the Section, we present the computation of the classical Noether charges. For our purposes the simplest worldline models are, see \cite{AKT1}, the ${\mathbb Z}_2\times{\mathbb Z}_2$-graded classical invariant actions of the 
$(1,2,1)_{[00]  }$ and $(1,2,1)_{[11]}$ multiplets. Both multiplets present one propagating bosonic field, two propagating fermionic fields and one auxiliary bosonic field. In the first multiplet the auxiliary field is the exotic boson, while in the second multiplet the auxiliary field is the ordinary boson (see \cite{AKT1} for details). The four time-dependent fields of respective ${\mathbb Z}_2\times{\mathbb Z}_2$
grading  ``$[00], [11], [10], [01]$" are accommodated into the multiplet $(x(t),z(t),\psi(t),\xi(t))^T$. In this paper it is more convenient to use the real time $t$, instead of the Euclidean time $\tau$ employed in \cite{AKT1}. Accordingly, the $D$-module representation acting on the $(1,2,1)_{[00]  }$ multiplet is defined by the operators
{{\bea\label{12100rep}
\qquad \qquad\qquad {\widehat  H}= \left(\begin{array}{cccc}i\partial_{t}&0&0&0\\0&i\partial_{t}&0&0\\0&0&i\partial_{t}&0\\ 
0&0&0&i\partial_{t}\end{array}\right), &\quad&
~~{\widehat Z}= \left(\begin{array}{cccc}0&1&0&0\\ \partial_{t}^2&0&0&0\\0&0&0&-i\partial_{t}
\\ 0&0&i\partial_{t}&0\end{array}\right),\nonumber\\
{\widehat Q}_{10}= \left(\begin{array}{cccc}0&0&1&0\\0&0&0&i\partial_{t}\\ i\partial_{t}&0&0&0\\ 
0&1&0&0\end{array}\right), &\quad&
{\widehat Q}_{01}= \left(\begin{array}{cccc}0&0&0&1\\0&0&-i\partial_t&0\\0&-1&0&0
\\ i\partial_{t}&0&0&0\end{array}\right).
\eea
}}
They close the ${\mathbb Z}_2\times {\mathbb Z}_2$-graded supertranslation algebra (\ref{z2z2super}) 
defined by the (anti)commutators
{{\bea\label{z2z2super2}
&\{{\widehat Q}_{10},{\widehat Q}_{10}\}=\{{\widehat Q}_{01},{\widehat Q}_{01}\}=2{\widehat H}, \quad [{\widehat Q}_{10},{\widehat Q}_{01}] = -2{\widehat Z},\quad [{\widehat H}, {\widehat Q}_{10}]=[{\widehat H},{\widehat Q}_{01}]=[{\widehat H},{\widehat Z}]=0.\quad&
\eea}}
The $D$-module representation acting on the $(1,2,1)_{[11]  }$ multiplet is defined by the operators
{{\bea\label{12111rep}
\qquad \qquad\qquad {\widecheck  H}= \left(\begin{array}{cccc}i\partial_{t}&0&0&0\\0&i\partial_{t}&0&0\\0&0&i\partial_{t}&0\\ 
0&0&0&i\partial_{t}\end{array}\right), &\quad&
~~{\widecheck Z}= \left(\begin{array}{cccc}0&- \partial_{t}^2&0&0\\ -1&0&0&0\\0&0&0&-i\partial_{t}
\\ 0&0&i\partial_{t}&0\end{array}\right),\nonumber\\
{\widecheck Q}_{10}= \left(\begin{array}{cccc}0&0&i\partial_t&0\\0&0&0&1\\ 1&0&0&0\\ 
0&i\partial_t&0&0\end{array}\right), &\quad&
{\widecheck Q}_{01}= \left(\begin{array}{cccc}0&0&0&i\partial_t\\0&0&-1&0\\0&-i\partial_t&0&0
\\ 1&0&0&0\end{array}\right).
\eea
}}
They close the (\ref{z2z2super2}) algebra (in the notation, the ``~$~{\widecheck{}}~$~" symbol replaces ``~$~{\widehat{}}~$~").\par
The field transformations are respectively read from (\ref{12100rep}) and (\ref{12111rep}). We have
\begin{equation}
  {\widehat{Q}}_{10}: 
  \begin{pmatrix}
     x \\ z \\ \psi \\ \xi
  \end{pmatrix}
  \mapsto 
  \begin{pmatrix}
 \psi \\ i\dot{\xi} \\ i \dot{x} \\ z
  \end{pmatrix},
   \qquad  
  {\widehat{Q}}_{01}: 
  \begin{pmatrix}
     x \\ z \\ \psi \\ \xi
  \end{pmatrix}
  \mapsto 
  \begin{pmatrix}
     \xi \\ -i\dot{\psi} \\ -z \\ i\dot{x}
  \end{pmatrix},
  \qquad
 {\widehat{ Z}}:
  \begin{pmatrix}
     x \\ z \\ \psi \\ \xi
  \end{pmatrix}
  \mapsto 
  \begin{pmatrix}
     z \\ \ddot{x} \\ -i \dot{\xi} \\ i\dot{\psi}
  \end{pmatrix}
  \label{QQZtransf00}
\end{equation}
and
\begin{equation}
  {\widecheck{Q}}_{10}: 
  \begin{pmatrix}
     x \\ z \\ \psi \\ \xi
  \end{pmatrix}
  \mapsto 
  \begin{pmatrix}

    i{\dot \psi} \\ {\xi} \\ {x} \\ i{\dot{ z}}
    
  \end{pmatrix},
   \qquad  
  {\widecheck{Q}}_{01}: 
  \begin{pmatrix}
     x \\ z \\ \psi \\ \xi
  \end{pmatrix}
  \mapsto 
  \begin{pmatrix}
    i{\dot{ \xi }}\\ -{\psi} \\ -i{\dot{z}} \\ {x}
  \end{pmatrix},
  \qquad
 {\widecheck{ Z}}:
  \begin{pmatrix}
     x \\ z \\ \psi \\ \xi
  \end{pmatrix}
  \mapsto 
  \begin{pmatrix}
     -{\ddot{z }}\\ -{x} \\ -i \dot{\xi} \\ i\dot{\psi}
  \end{pmatrix}.
  \label{QQZtransf11}
\end{equation}
The operator ${\widehat{H}}={\widecheck {H}}$ maps the fields into their time derivatives multiplied by $i$.\par
In the construction of the classical actions the $[10]$-graded and the $[01]$-graded component fields are assumed to be Grassmann. It is a consequence of the more general (\ref{z2z2brackets}) prescription for the (anti)commutators of the graded component fields. The action of the operators (\ref{12100rep}) and (\ref{12111rep}) on the graded component fields is assumed to satisfy the ${\mathbb Z}_2\times{\mathbb Z}_2$-graded Leibniz rule. \par
For the $(1,2,1)_{[00]}$ multiplet, the classical action ${\cal S} =\int dt {\cal L}$, invariant under the  (\ref{QQZtransf00}) transformations and (\ref{z2z2super2}) algebra, is given \cite{AKT1} by the Lagrangian
\bea\label{p00model}
  &{\cal L} = {\cal L}_\sigma + {\cal L}_{lin},&\nonumber\\&{\textrm{where}}\qquad 
  {\cal L}_\sigma = \frac{1}{2}\phi(x) (\dot{x}^2-z^2 + i\psi \dot{\psi} -i \xi \dot{\xi}) - 
  \frac{1}{2} \phi_x(x) z \psi \xi\qquad {\textrm{and}}\qquad
  {\cal L}_{lin} = \mu z. 
\eea

{\color{black}
We have denoted $ \phi_x = \frac{d \phi(x)}{dx}. $ 
The coupling constant $\mu$ has ${\mathbb Z}_2\times{\mathbb Z}_2$-grading $ \deg(\mu) = [11]$; this is required for the Lagrangian ${\cal L}_{\sigma}$ to be [00]-graded. $\mu$ can be interpreted as a classical, non-dynamical, background field, see \cite{AKT1}. We further mention that, in supersymmetric classical mechanics,  an odd-graded coupling constant was introduced in \cite{man}.
}
\par
The Lagrangian term ${\cal L}_\sigma$ is written, in manifestly invariant form, as
\begin{equation}
  {\cal L}_\sigma = -\frac{1}{2} {\widehat Z} {\widehat Q}_{10} {\widehat Q}_{01} g(x) + \frac{1}{2} \frac{d}{dt}(g_x \dot{x}), \qquad \phi(x) := g_{xx}(x).  \label{L0manifest}
\end{equation}
 The Euler-Lagrange equation for a component field $q$, which can be expressed, taking into account how fields are ordered, either as 
\bea  { \frac{d}{dt} }({\overrightarrow{\partial}}_{\dot{q}}{\cal L} ) -{ \overrightarrow{\partial}}_{q} {\cal L} &=& 0
\eea 
or as
\bea  { \frac{d}{dt} }({\cal L}{\overleftarrow{\partial}}_{\dot{q}} ) -{{\cal L} \overleftarrow{\partial}}_{q} &=& 0,
\eea 
produces in both cases the same set of equations:
\bea
 \phi \ddot{x} &=& -\frac{1}{2} \phi_x(\dot{x}^2 + z^2 -i \psi \dot{\psi}+i\xi \dot{\xi}) - \frac{1}{2} \phi_{xx} z \psi \xi,
 \nn \\
 2 \phi z &=& - \phi_x \psi \xi + 2 \mu,
 \nn \\
 i\phi \dot{\psi} &=& -\frac{1}{2} \phi_x (i \dot{x} \psi + z \xi),
 \nn \\
 i \phi \dot{\xi} &=& - \frac{1}{2} \phi_x (i \dot{x} \xi - z \psi). 
 \label{EL1}
\eea
For the $(1,2,1)_{[11]}$ multiplet the invariant classical action ${\cal S} =\int dt {\cal{\overline L}}$ is obtained, see \cite{AKT1}, from the Lagrangian
\bea\label{p11model}
  &{\cal {\overline L}} = {\cal {\overline L}}_\sigma + {\cal{\overline L}}_{lin},&
  \nn \\
  &{\textrm{where}}\qquad {\cal {\overline L}}_\sigma = \frac{1}{2}\Phi(z) (\dot{z}^2-x^2 + i\psi \dot{\psi} -i \xi \dot{\xi}) +  
  \frac{1}{2} \Phi_z(z) x \psi \xi\qquad {\textrm{and}}\qquad
  {\cal {\overline L}}_{lin} = {\overline \mu} x.&
\eea
As notable differences with the previous case $ \Phi(z)$ is an even function of $z$ and the coupling constant ${\overline \mu}$ is real, its ${\mathbb Z}_2\times{\mathbb Z}_2$ grading being given by $ \deg({\overline \mu}) = [00]$.\par

The Lagrangian term ${\overline{\cal L}}_\sigma$ is written, in manifestly invariant form, as
\begin{equation}
 {\overline {\cal L}}_\sigma = -\frac{1}{2} {\widecheck Z} {\widecheck Q}_{10} {\widecheck Q}_{01} f(z)+ \frac{1}{2} \frac{d}{dt}(f_z(z) \dot{z}), \qquad \Phi(z) := f_{zz}(z),  \label{L1manifest}
\end{equation}
with $f(z)$ an even function of $z$.

\par
The Euler-Lagrange equations of motion now read
\bea
 \Phi \ddot{z} &=& -\frac{1}{2} \Phi_z(\dot{z}^2 + x^2 -i \psi \dot{\psi}+i\xi \dot{\xi}) + \frac{1}{2} \Phi_{zz} x \psi \xi,
 \nn \\
 2 \Phi x &= & \Phi_z \psi \xi + 2{\overline \mu},
 \nn \\
 i\Phi \dot{\psi} &= &\frac{1}{2} \Phi_z (-i \dot{z} \psi + x \xi),
 \nn \\
 i \Phi \dot{\xi} &= &- \frac{1}{2} \Phi_z (i \dot{z} \xi + x \psi). 
 \label{EL6}
\eea

\subsection{The Noether charges}

We recall the general construction of the Noether charges. 
An action $
   S = \int  {\cal L}(q_i(t), \dot{q}_i(t)) dt
$ is invariant under the variation $
   q_i(t) \ \to \ q_i(t) + \delta q_i(t) \label{varq}
$ provided that there exists a $ \Lambda(t)$ such that
\begin{equation}
  \delta S = \int \dot{\Lambda}(t) dt.
\end{equation}
For our ${\mathbb Z}_2\times {\mathbb Z}_2$-graded fields the variation $ \delta S$ is computed as 
\bea
  \delta S &=& \int dt \big(   \delta q_i \pl{q_i} {\cal L} + \delta \dot{q}_i \pl{\delta\dot{q}_i} {\cal L} \big)= \int dt \left[ \delta q_i \Big(  \pl{q_i} {\cal L} - \frac{d}{dt}( \pl{\dot{q}_i} {\cal L}) \Big)   + \frac{d}{dt} \big(  \delta q_i \pl{\dot{q}_i} {\cal L} \big) \right].
\eea
By using the equations of motion the invariance of $S$ produces the identity
\begin{equation}
   \int dt \frac{d}{dt} \big( \Lambda - \delta q_i \pl{\dot{q}_i} {\cal L} \big) = 0
\end{equation}
which implies the existence of a conserved charge $Q$ given by
\begin{equation}
   Q = \Lambda - \delta q_i \pl{\dot{q}_i} {\cal L}. \label{Ncharge}
\end{equation}

Applying formula \eqref{Ncharge} to the Lagrangian \eqref{p00model} under a  \eqref{QQZtransf00} transformation we easily obtain $ \Lambda $.  Since no confusion arises, for convenience  we use the same notation for the generator of the transformation and its corresponding conserved charge. \\
The Noether charges for the invariant action of the $(1,2,1)_{[00]}$ multiplet are therefore given as 
\bea
  {\widehat H} &=& \frac{1}{2} \phi (\dot{x}^2 + z^2) + \frac{1}{2}\phi_x z \psi \xi - \mu z,
  \nn \\
  {\widehat Q}_{10} &=& \phi \dot{x} \psi + i \mu \xi,
  \nn \\
  {\widehat Q}_{01} &= &\phi \dot{x} \xi - i \mu \psi,
  \nn \\
  {\widehat Z} &=& \big( \phi  z + \frac{1}{2} \phi_x  \psi \xi - \mu \big) \dot{x}.
  \label{Ncharges00}
\eea
The following intermediate results were used to compute \eqref{Ncharges00}:
\begin{equation}
  \begin{array}{|c|c|c|}\hline
        & \Lambda & \delta_{q_i} \pl{\dot{q}_i} {\cal L} \\ \hline
      {\widehat H}: & 
      {\cal L} & 
      \displaystyle \phi \Big( \dot{x}^2 + \frac{i}{2} \psi \dot{\psi} - \frac{i}{2} \xi \dot{\xi} \Big)
      \\[10pt] \hline
      {\widehat Q}_{10}: &
      \displaystyle \frac{1}{2} \phi (iz \xi + \dot{x} \psi) - i \mu \xi & 
      \displaystyle  \frac{1}{2} \phi( 3 \dot{x} \psi + i z \xi) 
      \\[10pt]\hline
      {\widehat Q}_{01}: &
      \displaystyle \frac{1}{2} \phi (i z \psi - \dot{x} \xi) + i \mu \psi &
      \displaystyle  \frac{1}{2}\phi(\dot{x} \xi + i z \psi)
      \\[10pt]\hline
      {\widehat Z}: & 
      \displaystyle -\frac{1}{2} \Big( \phi_x \dot{x} \psi \xi + \phi \frac{d}{dt} (\psi \xi) \Big) + \mu \dot{x} &
      \displaystyle \phi \Big(  \dot{x}z - \frac{1}{2} \frac{d}{dt} (\psi \xi) \Big)\\[10pt]\hline
  \end{array}
\end{equation}
The Noether charges  $ {\widehat Q}_{10} $,  ${\widehat  Q}_{01}$ in (\ref{Ncharges00}) do not depend on the auxiliary field $z.$ \par
By using the algebraic relation for $z$ in the \eqref{EL1} equations of motion,  we can eliminate $z$ from all formulae obtained above. We get
\begin{equation}
 z = -\frac{\phi_x}{2\phi} \psi \xi + \frac{\mu}{\phi}.
\end{equation}
With this position the Lagrangian \eqref{p00model} now reads
\begin{equation}
   {\cal L} = \frac{1}{2}\phi (\dot{x}^2 + i \psi \dot{\psi} - i \xi \dot{\xi}) - \frac{\mu \phi_x}{2\phi} \psi \xi + \frac{\mu^2}{2\phi}.
\end{equation}
The Euler-Lagrange equations \eqref{EL1} become
\begin{align}
  \phi \ddot{x} &= - \frac{1}{2} \phi_x (\dot{x}^2 - i\psi \dot{\psi} + i \xi \dot{\xi} )
  - \frac{\mu}{2} \Big( \frac{\phi_x}{\phi} \Big)_x \psi \xi
  -\frac{\mu^2}{2} \frac{\phi_x}{\phi^2},
  \nn \\
  i \phi \dot{\psi} &= - \frac{1}{2} \phi_x \Big(  i\dot{x} \psi + \frac{\mu}{\phi} \xi \Big),
  \nn \\
  i \phi \dot{\xi} &= - \frac{1}{2} \phi_x \Big(  i\dot{x} \xi - \frac{\mu}{\phi} \psi \Big).  \label{EL2}
\end{align}
From the second and third equations we obtain
\begin{equation}
   i \phi \psi \dot{\psi} = - i \phi \xi \dot{\xi} = \mu \frac{\phi_x}{2 \phi}\psi \xi.
\end{equation}
These relations simplify the first equation in \eqref{EL2}; at the end we get the equations of motion
\bea
  \phi \ddot{x} &=& -\frac{1}{2} \phi_x \dot{x}^2 
    + \mu \Big( \frac{\phi_x^2}{\phi^2}  - \frac{\phi_{xx}}{2\phi} \Big) \psi \xi
    - \frac{\mu^2}{2} \frac{\phi_x}{\phi^2},
    \nn \\
  i \phi \dot{\psi} &=& - \frac{1}{2} \phi_x \Big(  i\dot{x} \psi + \frac{\mu}{\phi} \xi \Big),
  \nn \\
  i \phi \dot{\xi} &=& - \frac{1}{2} \phi_x \Big(  i\dot{x} \xi - \frac{\mu}{\phi} \psi \Big).  \label{EL3}
\eea
The Noether charges now read as
\bea
  {\widehat H} &=& \frac{1}{2} \phi \dot{x}^2 + \mu \frac{\phi_x}{2\phi} \psi \xi - \frac{\mu^2}{2\phi},
  \nn \\
  {\widehat Q}_{10} &=& \phi \dot{x} \psi + i \mu \xi,
  \nonumber\\
 {\widehat  Q}_{01} &=& \phi \dot{x} \xi - i \mu \psi,
  \nn \\
  {\widehat Z} &=& 0. \label{Ncharge00-2}  
\eea
One should note that the Noether charge ${\widehat  Z}$ now vanishes.\par
The (\ref{Ncharge00-2}) Noether charges are conserved under the  \eqref{EL3} equations of motion:
\begin{equation}
  \frac{d}{dt}{\widehat  H} = \frac{d}{dt}{\widehat  Q}_{10} = \frac{d}{dt} {\widehat Q}_{01} = 0.
\end{equation}
The same procedure can be repeated to compute the Noether charges of the $(1,2,1)_{[11]}$ model defined by the Lagrangian (\ref{p11model}).  The results corresponding to formula (\ref{Ncharges00}) in this case are
\bea
  {\widecheck H} &=& \frac{1}{2} \Phi (\dot{z}^2 + x^2) - \frac{1}{2}\Phi_z x \psi \xi -{\overline \mu} x,
  \nn \\
  {\widecheck Q}_{10} &=& \Phi \dot{z} \xi + i{\overline \mu }\psi,
  \nn \\
 {\widecheck  Q}_{01} &=& \Phi \dot{z} \psi - i{\overline \mu }\xi,
  \nn \\
 {\widecheck  Z} &=& \big( \Phi  x - \frac{1}{2} \Phi_z  \psi \xi - {\overline\mu} \big) \dot{z}.
  \label{Ncharges11}
\eea

\section{The Hamiltonian formulation of the classical models}

We discuss here the Hamiltonian formulation of the classical models introduced in Section {\bf 2}. In order to apply the canonical quantization of the models in Section {\bf 4}, following the procedure of  \cite{CHT} in connection with supersymmetric mechanics, 
we reexpress at first the component fields entering the given multiplet  in terms of a new basis called  the ``constant kinetic basis". We then introduce  the Hamiltonian dynamics, defined by Poisson and Dirac brackets, in this basis. To avoid unnecessary doubling of the text we extensively discuss the  $(1,2,1)_{[00]}$ model, while presenting only the main results concerning the $(1,2,1)_{[11]}$ model.

\subsection{Classical Lagrangians in the constant kinetic basis}

For the $(1,2,1)_{[00]}$ model the ``constant kinetic basis" which eliminates $\phi$ from the kinetic term of \eqref{p00model} is reached through the positions
\begin{equation}
  y = y(x), \qquad u = \sqrt{\phi(x)}\; z, \qquad \overline{\psi} = \sqrt{\phi(x)}\; \psi, \qquad \overline{\xi} = \sqrt{\phi(x)}\; \xi ,
  \label{newfield}
\end{equation}
where $y$ satisfies 
\bea
& y_x  = \sqrt{\phi(x)}, \qquad {\textrm{which implies}} \qquad \phi_x(y) = \phi_y(y) y_x = \sqrt{\phi(y)}\; \phi_y(y). &
\eea

\par
We rewrite the Lagrangian, equations of motions and Noether charges of Section {\bf 2} in terms of the new set of fields \eqref{newfield}. \par
The Lagrangian \eqref{p00model} now reads
\begin{equation}
  {\cal L} = \frac{1}{2}( \dot{y}^2 - u^2 + i \overline{\psi}\, \dot{\overline{\psi}} - i \overline{\xi}\, \dot{\overline{\xi}} ) - \frac{1}{2} \frac{\phi_y}{\phi}\, u \overline{\psi} \,\overline{\xi} + \frac{\mu }{\sqrt{\phi}}\, u. 
  \label{p00model2}
\end{equation}
In terms of the new function $ W(y)$, introduced through
\begin{equation}\label{Wterm}
  W(y) := \frac{1}{\sqrt{\phi(y)}},
\end{equation}
we get
\begin{equation}
  {\cal L} = \frac{1}{2} (\dot{y}^2 - u^2 + i{\overline \psi} \dot{{\overline\psi}} - i{\overline \xi} \dot{{\overline \xi}}) + (\ln W)_y \, u {\overline \psi} {\overline \xi} + \mu W u.
\end{equation}

\noindent
The Euler-Lagrange equations \eqref{EL1} are written as
\begin{alignat}{2}
  \ddot{y} &= (\ln W)_{yy}\, u {\overline\psi}{\overline \xi} + \mu W_y\, u, 
  & \qquad 
  u &= (\ln W)_y {\overline \psi}{\overline  \xi} + \mu W,
  \nn \\
  i \dot{{\overline\psi}} &= (\ln W)_y\,u {\overline \xi}, 
  &
  i \dot{{\overline \xi}} &= -(\ln W)_y\, u {\overline \psi}. \label{EL4}
\end{alignat}
The Noether charges \eqref{Ncharges00} now read 
\bea\label{newnoetherwithW}
   {\widehat H} &=& \frac{1}{2}( \dot{y}^2 + u^2) - (\ln W)_y\,u {\overline\psi}{\overline \xi} - \mu W u,
   \nn \\
   {\widehat Q}_{10} &=& \dot{y} {\overline\psi} + i \mu W {\overline\xi}, 
 \nn \\
   {\widehat Q}_{01} &= &\dot{y}{\overline \xi} - i \mu W{\overline \psi},
   \nn \\
   {\widehat Z} &= &\dot{y} (u - (\ln W)_y {\overline \psi}{\overline  \xi} - \mu W).
\eea
When eliminating the auxiliary field $u$ we get the Lagrangian
\begin{equation}
  {\cal L} = \frac{1}{2} (\dot{y}^2  + i {\overline\psi} \dot{{\overline\psi}} - i{\overline \xi} \dot{{\overline\xi}}) + \mu W_y {\overline{\psi}} {\overline \xi} + \frac{1}{2} \mu^2 W^2,
  \label{p00model3}
\end{equation}
the equations of motion
\bea
   \ddot{y} &=& \mu W_{yy} {\overline\psi}{\overline \xi} + \mu^2 W_y W, 
   \nn \\
   i\dot{{\overline{\psi}}} &= &\mu W_y {\overline\xi}, \nonumber\\
   i\dot{{\overline\xi}}& =& - \mu W_y {\overline\psi}
   \label{EL5}
\eea
and the Noether charges
\bea
 {\widehat  H} &=& \frac{1}{2} \dot{y}^2 - \mu W_y{\overline \psi}{\overline \xi} - \frac{1}{2} \mu^2 W^2,
  \nn \\
 {\widehat  Q}_{10} &=& \dot{y}{\overline \psi} + i \mu W {\overline\xi}, \nn \\
 {\widehat  Q}_{01} &=& \dot{y} {\overline\xi} - i\mu W {\overline\psi},
  \nn \\
  {\widehat Z} &= &0.
  \label{Ncharge00-3}
\eea
For the $(1,2,1)_{[11]}$ model the constant kinetic basis is given by
\begin{equation}
  \tilde{x} = \sqrt{\Phi(z)} x, \quad
  \tilde{z} = C(z), \quad
  \tilde{\psi} = \sqrt{\Phi(z)} \psi, \quad
  \tilde{\xi} = \sqrt{\Phi(z)} \xi,    
\end{equation}
where $ C(z)$ is a  function with ${\mathbb Z}_2\times{\mathbb Z}_2$-grading $[11]$. It satisfies the relation
\bea
&C_z(z) = \sqrt{\Phi(z)}, \qquad {\textrm{so that}}\qquad
   \dot{\tilde{z}}^2 = \dot{z}^2 \Phi(z), \qquad
   \Phi_z(z) = \sqrt{\Phi(\tilde{z}) }\, \Phi_{\tilde{z}}(\tilde{z}).&
\eea
The consistency requires that $\sqrt{\Phi(z)}$ is an even, $[00]$-graded function. \par
After defining ${\overline W}(z)$ to be
\begin{equation}
    {\overline W}(z) := \frac{{\overline\mu}}{\sqrt{\Phi(z)}},
\end{equation}
by repeating the same steps as before the Lagrangian \eqref{p11model} is expressed in the new basis as 
\begin{equation}
  {\cal {\overline L}} = \frac{1}{2} (\dot{{\tilde z}}^2 - {\tilde x}^2 + i{\tilde \psi} \dot{{\tilde\psi}} - i {\tilde\xi}\dot{{\tilde \xi}}) - (\ln {\overline W}({\tilde z}))_{\tilde z} {\tilde x} {\tilde \psi} {\tilde \xi}  +{\overline\mu} {\overline W}({\tilde z}){\tilde x}.
  \label{redefL211}
\end{equation}
After eliminating the auxiliary field ${\tilde x}$,
the Lagrangian becomes
\begin{equation}
  {\cal {\overline L}} = \frac{1}{2} (\dot{{\tilde z}}^2  + i{\tilde \psi} \dot{{\tilde \psi}} - i{\tilde \xi}\dot{{\tilde \xi}})  -  {\overline W}_{\tilde z}({\tilde z}) {\tilde \psi}{\tilde \xi} + \frac{1}{2}{\overline  W}({\tilde z})^2,
  \label{redefL311}
\end{equation}
while its associated Noether charges are
\bea
   {\widecheck H} &=& \frac{1}{2} \dot{{\tilde z}}^2  +   {\overline W}_{\tilde z}({\tilde z}) {\tilde \psi} {\tilde \xi} - \frac{1}{2}  {\overline W}({\tilde z})^2,
   \nonumber \\
   {\widecheck Q}_{10} &=& \dot{{\tilde z}} {\tilde \xi} + i {\overline  W}({\tilde z}) {\tilde \psi}, 
   \nonumber \\
   {\widecheck Q}_{01} &= &\dot{{\tilde z}}{\tilde  \psi} -i {\overline  W}({\tilde z}) {\tilde \xi},
   \nonumber \\
   {\widecheck Z} &=& 0.    \label{11charges3}
\eea

\subsection{The Hamiltonian mechanics of the $(1,2,1)_{[00]}$ model}

We introduce now the Hamiltonian formulation of the $(1,2,1)_{[00]}$ model.\par
The conjugate momenta computed from the Lagrangian \eqref{p00model3} are introduced through
\begin{equation}
   p_y = {\cal L} {\overleftarrow{\partial}}_{\dot{y}} = \dot{y}, \qquad
   p_{{\overline\psi}} =  {\cal L} {\overleftarrow{\partial}}_{\dot{{\overline\psi}}} = \frac{i}{2}{\overline\psi}, \qquad
   p_{{\overline\xi}} =  {\cal L} {\overleftarrow{\partial}}_{\dot{{\overline\xi}}} = -\frac{i}{2}{\overline\xi}. 
   \label{momenta}
\end{equation}
The Hamiltonian $H$ is defined as
\begin{equation}\label{canonhamil}
  H := p_{q_i} \dot{q}_i - {\cal L} = \frac{1}{2} p_y^2 - \mu W_y {\overline \psi }{\overline \xi} - \frac{1}{2} \mu^2 W^2. 
\end{equation}
One should note that the Hamiltonian is identical to the Noether charge ${\widehat H}$ given in \eqref{Ncharge00-3}. \par
In terms of the momenta \eqref{momenta} the Noether charges are
\begin{equation}
  {\widehat H} = \frac{1}{2} p_y^2 - \mu W_y {\overline \psi}{\overline \xi} - \frac{1}{2} \mu^2 W^2, 
  \quad
  {\widehat Q}_{10} = p_y {\overline\psi} + i \mu W {\overline \xi}, \qquad
  {\widehat Q}_{01} = p_y {\overline\xi} - i\mu W {\overline\psi}.
  \label{Ncharge00-4}
\end{equation}
We have two constraints, given by
\begin{equation}
  f_1 = p_{{\overline\psi}} - \frac{i}{2}{\overline\psi}, \qquad f_2 = p_{{\overline\xi}}+ \frac{i}{2}{\overline\xi},\label{f12}
\end{equation}
whose ${\mathbb Z}_2\times {\mathbb Z}_2$ grading are $ \deg(f_1) = [10]$, $\deg(f_2) = [01]$. \par
The Poisson brackets are conveniently introduced through
\begin{equation}
  \PB{A,B} = A \widehat{\Gamma} B - (-1)^{{\bf{a}}\cdot {\bf{b}}} B \widehat{\Gamma}A, \qquad
  \widehat{\Gamma} = {\overleftarrow{\partial}}_{y} {\overrightarrow{\partial}}_{p_y} +{\overleftarrow{\partial}}_{{\overline{\psi}}} {\overrightarrow{\partial}}_{p_{\overline{\psi}}}  +{\overleftarrow{\partial}}_{{\overline{\xi}}} {\overrightarrow{\partial}}_{p_{\overline{\xi}}},
  \label{PB00def}  
\end{equation}
where $ \deg(A) = {\bf{a}}, \; \deg(B) ={ \bf{b}}.$ \par
One may easily see that the constraints  \eqref{f12} are of second class:
\begin{equation}
   \PB{f_1, f_1} = - \PB{f_2, f_2} = -i, \qquad
   \PB{f_1, f_2} = 0.
\end{equation}
The computation uses the identities
\begin{alignat}{2}
  f_1 
  \begin{pmatrix}
    \pr{y} \\ \pr{\psi} \\ \pr{\xi}
  \end{pmatrix}
  &= 
  \begin{pmatrix}
    0 \\ -i/2 \\ 0
  \end{pmatrix},
  & \qquad
  f_2 
  \begin{pmatrix}
    \pr{y} \\ \pr{\psi} \\ \pr{\xi}
  \end{pmatrix}
  &= 
  \begin{pmatrix}
    0 \\ 0 \\ i/2
  \end{pmatrix},
  \nn  \\
  \begin{pmatrix}
    \pl{p_y} \\ \pl{p_{\psi}} \\ \pl{p_{\xi}}
  \end{pmatrix}
  f_1
  &=
  \begin{pmatrix}
     0 \\ 1 \\ 0
  \end{pmatrix},
  &
  \begin{pmatrix}
    \pl{p_y} \\ \pl{p_{\psi}} \\ \pl{p_{\xi}}
  \end{pmatrix}
  f_2
  &=
  \begin{pmatrix}
     0 \\ 0 \\ 1
  \end{pmatrix}.  
\end{alignat}
As easily seen from the definition \eqref{PB00def}, the
nonvanishing Poisson brackets of the canonical variables are
\begin{equation}
  \PB{y,p_y} = \PB{\psi, p_{\psi}} = \PB{\xi, p_{\xi}} = 1. \label{PBcanonicalv}
\end{equation}
The Poisson brackets of the Noether charges \eqref{Ncharge00-4} are computed by using the relations
\begin{align}
  {\widehat Q}_{01}
  \begin{pmatrix}
    \pr{y} \\ \pr{{\overline\psi}} \\ \pr{{\overline\xi}}
  \end{pmatrix}
  &= 
  \begin{pmatrix}
    -i \mu W_y {\overline\psi} \\ -i \mu W \\ p_y
  \end{pmatrix},
   \qquad
  {\widehat Q}_{10}
  \begin{pmatrix}
    \pr{y} \\ \pr{{\overline\psi}} \\ \pr{{\overline\xi}}
  \end{pmatrix}
  = 
  \begin{pmatrix}
    i \mu W_y {\overline\xi} \\ p_y \\ i \mu W 
  \end{pmatrix},
  \nn \\
  {\widehat H}
  \begin{pmatrix}
    \pr{y} \\ \pr{{\overline\psi}} \\ \pr{{\overline\xi}}
  \end{pmatrix}
  &= 
  \begin{pmatrix}
    -\mu W_{yy} {\overline\psi}{\overline \xi} - \mu^2 W_y W 
    \\
    -\mu W_y {\overline\xi }
    \\
    -\mu W_y {\overline\psi}
  \end{pmatrix}
\end{align}
and
\begin{equation}
  \begin{pmatrix}
    \pl{p_y} \\ \pl{p_{{\overline\psi}}} \\ \pl{p_{{\overline\xi}}}
  \end{pmatrix}
  {\widehat Q}_{01}
  =
  \begin{pmatrix}
    {\overline \xi} \\ 0 \\ 0
  \end{pmatrix},
  \qquad
  \begin{pmatrix}
    \pl{p_y} \\ \pl{p_{{\overline\psi}}} \\ \pl{p_{{\overline\xi}}}
  \end{pmatrix}
  {\widehat Q}_{10}
  =
  \begin{pmatrix}
    {\overline \psi}
 \\ 0 \\ 0
  \end{pmatrix},  
  \qquad 
  \begin{pmatrix}
    \pl{p_y} \\ \pl{p_{{\overline\psi}}} \\ \pl{p_{{\overline\xi}}}
  \end{pmatrix}
  {\widehat H} =
  \begin{pmatrix}
     p_y \\ 0 \\ 0
  \end{pmatrix}.
\end{equation}
One proves that the Poisson bracket of the pair  $ {\widehat Q}_{01},{\widehat  Q}_{10}$ vanishes
in accordance with the ${\widehat  Z} = 0$ relation (\ref{Ncharge00-2}).\par All Poisson brackets among the Noether charges \eqref{Ncharge00-4} are 
\begin{align}
   \PB{{\widehat Q}_{01}, {\widehat Q}_{01}} &= -\PB{{\widehat Q}_{10},{\widehat  Q}_{10}} = -2i \mu W_y{\overline \psi }{\overline \xi},
   \nn \\
   \PB{{\widehat Q}_{01},{\widehat  Q}_{10}} &= 0,
   \nn \\
   \PB{{\widehat Q}_{01},{\widehat H}} &= -i \mu W_y {\overline \psi} p_y + \mu^2 W_y W{\overline  \xi},
   \nn \\
   \PB{{\widehat Q}_{10},{\widehat H}} &= i \mu W_y {\overline \xi} p_y + \mu^2 W_y W{\overline  \psi}.
\end{align}
At the level of the Poisson bracket, the Noether charges \eqref{Ncharge00-4} do not recover the $\mathbb{Z}_2\times{\mathbb Z}_2$-graded supertranslation algebra (\ref{z2z2super2}).\par
Due to the presence of the second class constraints (\ref{f12}), the Dirac brackets are defined as
\begin{align}
  \DB{A,B} &= \PB{A,B} - \PB{A,f_i} (\Delta^{-1})_{ij} \PB{f_j, B}=
  \nn \\
  &=\PB{A,B} - i \PB{A,f_1} \PB{f_1,B} + i \PB{A,f_2} \PB{f_2,B},
  \label{DB00def}
\end{align}
where
\begin{equation}
  \Delta :=
  \begin{pmatrix}
    \PB{f_1, f_1} & \PB{f_1, f_2} \\[3pt]
    \PB{f_2, f_1} & \PB{f_2, f_2}
  \end{pmatrix}
  =
  \begin{pmatrix}
    -i & 0 \\[3pt] 0 & i
  \end{pmatrix},
  \qquad
  \Delta^{-1} = 
  \begin{pmatrix}
    i & 0 \\[3pt] 0 & -i
  \end{pmatrix}.
\end{equation}
  The Dirac brackets of the Noether charges \eqref{Ncharge00-4} satisfy, with ${\widehat Z}=0$, the $\mathbb{Z}_2\times{\mathbb Z}_2$-graded supertranslation algebra  (\ref{z2z2super2}):
\begin{align}
  \DB{{\widehat Q}_{01},{\widehat Q}_{01}} &= 2 i{\widehat  H}, \qquad \DB{{\widehat Q}_{10},{\widehat  Q}_{10}} = -2i{\widehat H},
  \nn \\
  \DB{{\widehat Q}_{01},{\widehat  Q}_{10}} &= \DB{{\widehat Q}_{01},{\widehat  H}} = \DB{{\widehat Q}_{10},{\widehat H}} = 0. 
  \label{DBclassicalalg}
\end{align}
This set of equation is verified by using the identities
\begin{alignat}{3}
 \PB{{\widehat Q}_{01},f_1} &= -i\mu W, & \qquad 
 \PB{{\widehat Q}_{01},f_2} &= p_y, & \qquad 
 \PB{{\widehat Q}_{10},f_1} &= p_y, 
 \nn \\
 \PB{{\widehat Q}_{10},f_2} &= i \mu W, & 
 \PB{f_1,{\widehat H}} &= \mu W_y {\overline \xi}, & 
 \PB{f_2, {\widehat H}} &= \mu W_y {\overline\psi}.
\end{alignat}
The canonical equations of motion obtained by the Dirac brackets with the Hamiltonian $H={\widehat H}$ of \eqref{Ncharge00-4} are identical to the Euler-Lagrange equations \eqref{EL5}:
\begin{alignat}{2}
  \dot{y} &= \DB{y,{\widehat  H}} = p_y, &\qquad 
  \dot{p_y} &= \DB{p_y, {\widehat H}} = \mu W_{yy} {\overline\psi}{\overline \xi} + \mu^2 W_y W,
  \nn \\
  \dot{{\overline\psi}} &= \DB{\psi,{\widehat  H}} = -i \mu W_y {\overline \xi}, &
  \dot{{\overline\xi}} &= \DB{{\overline\xi},{\widehat  H}} = i \mu W_y {\overline \psi}.
\end{alignat}
The nonvanishing Dirac bracket for the canonical variables are
\begin{equation}
  \DB{y, p_y} = 1, \qquad
  \DB{{\overline \psi}, {\overline\psi}} = -i, \qquad
  \DB{{\overline\xi},{\overline \xi}} = i. 
  \label{DBcanonicalv}
\end{equation}
This is seen as follows: let $ \deg(A) = {\bf{a}} = [a_1,a_2]$; then  one immediately checks that 
\begin{equation}
  \PB{A, f_1} = A \pr{{\overline \psi}} + (-1)^{a_1} \frac{i}{2} \pl{p_{{\overline\psi}}} A
\end{equation}
implies that $ f_1$ has nonvanishing Poisson brackets only with $ {\overline\psi} $ and $ p_{{\overline\psi}}$,
\begin{equation}
  \PB{{\overline \psi}, f_1} = 1, \qquad 
  \PB{p_{{\overline \psi}}, f_1} = - \frac{i}{2}.
\end{equation}
Similarly,
\begin{equation}
  \PB{A, f_2} = A \pr{{\overline \xi}} - (-1)^{a_2} \frac{i}{2} \pl{p_{{\overline \xi}}} A
\end{equation}
implies that $ f_2$ has nonvanishing Poisson brackets only with ${\overline \xi} $ and $ p_{{\overline\xi}}:$
\begin{equation}
  \PB{{\overline \xi}, f_2} = 1, \qquad 
  \PB{p_{{\overline \xi}}, f_2} = \frac{i}{2}.
\end{equation}
Combining these nonvanishing Poisson brackets with the ones given in \eqref{PBcanonicalv}, it is not difficult to see that, in addition to \eqref{DBcanonicalv}, further nonvanishing Dirac brackets for the canonical variables are
\begin{equation}
  \DB{{\overline \psi}, p_{{\overline\psi}}} = \DB{{\overline \xi}, p_{{\overline \xi}}} = \frac{1}{2}. 
\end{equation}
They are reduced to the ones in \eqref{DBcanonicalv} by using the expression of the momenta \eqref{momenta}.

\subsection{The Hamiltonian mechanics of the $(1,2,1)_{[11]}$ model}

We collect here the relevant formulas of the Hamiltonian formulation of the $(1,2,1)_{[11]}$ model. They are derived
from the Lagrangian mechanics  (\ref{redefL311}) by using the same procedure discussed in the previous subsection.\par
The conjugate momenta are
\begin{equation}
   p_{\tilde z} = {\overline{\cal L}} \pr{\dot{{\tilde z}}} = \dot{{\tilde z}}, \quad
   p_{{\tilde\psi}} ={\overline {\cal L}} \pr{\dot{{\tilde\psi}}} = \frac{i}{2} {\tilde \psi}, \quad
   p_{{\tilde\xi}} ={\overline {\cal L}} \pr{\dot{{\tilde\xi}}} = -\frac{i}{2}{\tilde \xi}.
\end{equation}
The
Hamiltonian $H={\widecheck H}$ is
\begin{equation}
   H ={\widecheck H}= \frac{1}{2} p_{\tilde z}^2 +  {\overline W}_{\tilde z}({\tilde z}) {\tilde\psi}{\tilde \xi }- \frac{1}{2}  {\overline W}({\tilde z})^2.
\end{equation}
It coincides with the Nother charge \eqref{11charges3}. 
The other nonvanishing Noether charges are
\begin{equation}
   {\widecheck Q}_{10} = p_{\tilde z} {\tilde\xi}  + i{\overline  W}({\tilde z}) {\tilde \psi}, \qquad {\widecheck Q}_{01} = p_{\tilde z}{\tilde \psi} - i {\overline W}({\tilde z}) {\tilde \xi}. 
   \label{NchargeHamilton}
\end{equation}
We present the Dirac brackets for the canonical variables. They are given by
\begin{equation}\label{db11model}
  \DB{{\tilde z}, p_{\tilde z}} = 1, \qquad \DB{{\tilde \psi}, {\tilde \psi}} = -i, \qquad \DB{{\tilde \xi},{\tilde \xi}} = i.
\end{equation}
The nonvanishing Dirac brackets of the Noether charges are
\begin{equation}
  \DB{{\widecheck Q}_{10},{\widecheck Q}_{10}} = -2i{\widecheck H}, \qquad \DB{{\widecheck Q}_{01}, {\widecheck Q}_{01}} = 2i{\widecheck H}.
\end{equation}
The canonical equations of motion defined by the Dirac brackets are
\begin{align}
   \dot{{\tilde z}} &= \DB{{\tilde z},{\widecheck H}} = p_{\tilde z},\nn
   \\
   \dot{p}_{\tilde z} &= \DB{p_{\tilde z}, {\widecheck H}} = - {\overline W}_{{\tilde z}{\tilde z} }{\tilde \psi}{\tilde  \xi} + {\overline W}_{\tilde z}{\overline  W},\nn
   \\
   \dot{{\tilde \psi}} &= \DB{{\tilde \psi},{\widecheck  H}} = i {\overline W}_{\tilde z} {\tilde \xi}, \nn
   \\
   \dot{{\tilde \xi}} &= \DB{{\tilde \xi},{\widecheck  H}} = -i{\overline  W}_{\tilde z} {\tilde \psi}.
\end{align}
They are  identical to the Euler-Lagrange equations obtained from \eqref{redefL311}. 

\section{The canonical quantization}

We present the canonical quantization of the $(1,2,1)_{[00]}$ and $(1,2,1)_{[11]}$ classical models.\par
As is customary, the quantization is obtained by replacing the Dirac brackets in \eqref{DBcanonicalv} and (\ref{db11model}) with (anti)commutators. It is obtained through the mapping
\bea
  \DB{. ,  . }\ &\mapsto&  -i [  ., . \},
\eea
where the ``$ [ ., . \} $" symbol introduced in (\ref{z2z2brackets})  denotes the $\mathbb{Z}_2\times{\mathbb Z}_2$-graded Lie brackets. 
\subsection{The quantization of the $(1,2,1)_{[00]}$ model}
For  the $(1,2,1)_{[00]}$ model the canonical Dirac brackets (\ref{DBcanonicalv}) are replaced by the (anti)commutators
\begin{equation}
  [y, p_y] = i, \qquad \{ {\overline \psi}, {\overline \psi} \} = 1, \qquad \{ {\overline \xi}, {\overline \xi} \} = -1, \qquad
  [{\overline \psi}, {\overline \xi}] = 0. 
  \label{CCR1}
\end{equation}

In the Heisenberg picture the (anti)commutators (\ref{CCR1}) are computed at equal time, let's say at $t=0$. A consistency condition further requires the $deg(\mu)=[11]$ coupling constant $\mu$ entering (\ref{canonhamil}) to anticommute with
${\overline\psi}$, ${\overline \xi}$, so that
\begin{equation}
  \{ \mu,{\overline \psi} \} = \{ \mu, {\overline \xi} \} = 0. \label{CCR2}
\end{equation}
It follows from \eqref{CCR1} and \eqref{CCR2} that the quantization of the Noether charges \eqref{Ncharge00-4} satisfy the relations:
\begin{align}
  \{ {\widehat Q}_{10},{\widehat  Q}_{10} \} &= 2{\widehat H}, \qquad 
  \{ {\widehat Q}_{01},{\widehat  Q}_{01} \} = -2{\widehat H},
  \nn \\
  [ {\widehat Q}_{01},{\widehat  Q}_{10} ] &= [{\widehat  Q}_{01},{\widehat  H}] = [ {\widehat Q}_{10},{\widehat H}] = 0.
\end{align}
Before going ahead we introduce the representations of the relations  (\ref{CCR1}) and  (\ref{CCR2}). \par
The quantum operators
${\overline\psi},{\overline\xi},\mu$ can be represented by $4\times 4$ matrices satisfying (\ref{gradedmatrices}), so that
\bea\label{gradedsolutions}
& {\overline \psi}\in {\cal G}_{10},\quad {\overline \xi}\in {\cal G}_{01},\quad \mu \in {\cal G}_{11}&
\eea
and
\bea\label{anticommutatorrelations}
&\{{\overline \psi},{\overline\psi}\}=-\{{\overline\xi},{\overline\xi}\}={\mathbb I}_4,\quad [{\overline \psi},{\overline \xi}]=0,\quad \{{\overline \psi},\mu\}=\{{\overline \xi},\mu\}=0.&
\eea
We are looking for real matrix solutions to the above system of equations.\par
This means that in particular ${\overline\psi}$ should be given by a linear combination $${\overline \psi}=\lambda_1 \cdot Y\otimes I+\lambda_2 \cdot Y\otimes X$$ of the split-quaternion matrices $I,X,Y,A$ introduced in (\ref{splitquat}). Similar relations hold for ${\overline\xi}$ and $\mu$.  The requirement that 
${\overline \psi}^2$ is proportional to the identity implies that $\lambda_{1,2}$ cannot be both different from $0$. Therefore, up to an overall sign, there are only two possible solutions
for ${\overline \psi}$ (a similar argument applies to ${\overline\xi}$ as well). The system of equations (\ref{anticommutatorrelations}) is solved (up to an overall sign for each one of the three matrices) by the two sets of triples:
\bea
&{\overline \psi}_A= \frac{1}{\sqrt 2} Y\otimes I,\quad {\overline \xi}_A= \frac{1}{\sqrt 2} Y\otimes A,\quad {\overline \mu}_A= X\otimes A& 
\eea
and
\bea
&{\overline \psi}_B= \frac{1}{\sqrt 2} Y\otimes X,\quad {\overline \xi}_B= \frac{1}{\sqrt 2} A\otimes Y,\quad {\overline \mu}_B= -I\otimes A.& 
\eea
It is convenient to introduce the triples of Hermitian matrices
\bea\label{triplesherm}
{\psi}_A={\overline \psi}_A,\quad { \xi}_A=i {\overline \xi}_A,\quad { \mu}_A = i{\overline \mu}_A\quad &{\textrm{and}}&
{\psi}_B={\overline \psi}_B,\quad {\xi}_B=i {\overline \xi}_B,\quad {\mu}_B = i{\overline \mu}_B,
\eea
so that
\bea
&{\psi}_{A,B}={\psi}_{A,B}^\dagger,\quad{\xi}_{A,B}={\xi}_{A,B}^\dagger,\quad
{\mu}_{A,B}={\mu}_{A,B}^\dagger.&
\eea
In terms of these representations and up to an overall normalization factor, the quantized Noether charges ${\widehat Q}_{10}, {\widehat Q}_{01}$ are associated to the hermitian conserved charges $Q_{10}^A$, $Q_{01}^A$ and  $Q_{10}^B$, $Q_{01}^B$, given by
\bea\label{fournoether}
Q_{10}^A= -i({\psi}_A\cdot\partial_y +W(y){\mu}_A{\xi}_A),&&
Q_{10}^B= -i({\psi}_B\cdot\partial_y +W(y){\mu}_B{\xi}_B),\nonumber\\
Q_{01}^A= -i({\xi}_A\cdot\partial_y +W(y){\mu}_A{\psi}_A),&&
Q_{10}^B= -i({\xi}_B\cdot\partial_y +W(y){\mu}_B{\psi}_B).
\eea
By construction the hermiticity conditions hold:
\bea
&
(Q_{10}^A)^\dagger=Q_{10}^A,\quad 
(Q_{10}^B)^\dagger=Q_{10}^B,\quad 
(Q_{01}^A)^\dagger=Q_{01}^A,\quad 
(Q_{01}^B)^\dagger=Q_{01}^B.&
\eea
These four supercharges are
\bea\label{absupercharges}
Q_{10}^A&=&-\frac{i}{\sqrt{2}} \left(\begin{array}{cccc}0&0&\partial_y+W(y)&0\\ 0&0&0&\partial_y+W(y)\\ \partial_y-W(y)&0&0&0\\ 0&\partial_y-W(y)&0&0\end{array}\right),\nonumber\\
Q_{10}^B&=&-\frac{i}{\sqrt{2}} \left(\begin{array}{cccc}0&0&\partial_y+W(y)&0\\ 0&0&0&-\partial_y-W(y)\\ \partial_y-W(y)&0&0&0\\ 0&-\partial_y+W(y)&0&0\end{array}\right),\nonumber\\
Q_{01}^A&=&\frac{1}{\sqrt{2}} \left(\begin{array}{cccc}0&0&0&\partial_y+W(y)\\ 0&0&-\partial_y-W(y)&0\\0& \partial_y-W(y)&0&0\\ -\partial_y+W(y)&0&0&0\end{array}\right),\nonumber\\
Q_{01}^B&=&\frac{1}{\sqrt{2}} \left(\begin{array}{cccc}0&0&0&\partial_y+W(y)\\ 0&0&\partial_y+W(y)&0\\ 0&-\partial_y+W(y)&0&0\\ -\partial_y+W(y)&0&0&0\end{array}\right).
\eea
The supercharges are the square roots of the Hamiltonian $H$, since
\bea\label{squarerootsof}
&\{Q_{10}^A,Q_{10}^A\}= \{Q_{10}^B,Q_{10}^B\}= \{Q_{01}^A,Q_{01}^A\}= \{Q_{01}^B,Q_{01}^B\}= 2H,&
\eea
where
\bea
H&=&H^\dagger
\eea
and
\bea\label{qham12100}
H&=&\frac{1}{2} \left(\begin{array}{cccc}-\partial_y^2+W^2+W'&0&0&0\\ 0&-\partial_y^2+W^2+W'&0&0\\ 0&0&-\partial_y^2+W^2-W'&0\\ 0&0&0&-\partial_y^2+W^2-W'\end{array}\right).\nonumber\\&&
\eea
In the above formula $W\equiv W(y)$ and $W'\equiv \frac{d}{dy}W(y)$.\par
One should note that, unlike the supercharges (\ref{fournoether}), the Hamiltonian $H$ does not depend on which choice of three matrices, either  $
{\psi}_A,~ { \xi}_A,~{ \mu}_A$ or ${\psi}_B,~ { \xi}_B,~{ \mu}_B$, is made.\par
The Hamiltonian (\ref{qham12100}) reproduces, up to a normalization convention and the reordering  of the diagonal elements by a similarity transformation, the ${\mathbb Z}_2\times{\mathbb Z}_2$-graded quantum Hamiltonian introduced in formula (3.1.1) of
\cite{BruDup}. As promised, we obtained this quantum model from the canonical quantization of the ${\mathbb Z}_2\times{\mathbb Z}_2$-graded classical model based on the
$(1,2,1)_{[00]}$ multiplet.\par
{\color{black}
The spectrum of the Hamiltonian (\ref{qham12100}) in the $\mathbb{Z}_2\times\mathbb{Z}_2$-graded Hilbert space was investigated in \cite{BruDup}; the ground state and the excited states are two-fold and four-fold degenerate, respectively. }\par
We point out that the four matrices $\psi_A, \psi_B,\xi_A, \xi_B$ from (\ref{triplesherm}) do not commute with the Hamiltonian. This is in agreement with the fact that they correspond to the quantization of
the classical dynamical variables ${\overline\psi}$, ${\overline\xi}$ and that their Heisenberg evolution is expected. The matrices $\mu_A$, $\mu_B$ from (\ref{triplesherm}) are, in their respective representations, the quantum counterparts of the classical coupling constant $\mu$. It is rewarding that they commute with the Hamiltonian:
\bea
&[H,\mu_A]=[H,\mu_B]=0.&
\eea
This is an extra consistency check of the correctness of the proposed quantization prescription.  All quantized
coupling constants introduced in the following commute with the respective Hamiltonians
{\color{black}
and act on a $\mathbb{Z}_2 \times \mathbb{Z}_2$-graded Hilbert space as ordinary constant matrices. } 

\subsection{The quantization of the $(1,2,1)_{[11]}$ model}

For  the $(1,2,1)_{[11]}$ model the  canonical Dirac brackets (\ref{db11model}) are replaced by the (anti)commutators
\begin{equation}
  [{\tilde z}, p_{\tilde z}] = i, \qquad \{ {\tilde\psi}, {\tilde \psi} \} = 1, \qquad \{ {\tilde \xi}, {\tilde \xi} \} = -1, \qquad
  [{\tilde \psi}, {\tilde \xi}] = 0. 
  \label{CCR1case11}
\end{equation}
The coupling constant ${\tilde \mu} $ has now $\deg({\tilde\mu})=[00]$, so that
\bea\label{ccr1extra1}
&[{\tilde \psi},{\tilde\mu}]=[{\tilde\xi},{\tilde\mu}]=0.&
\eea
On the other hand, since $\deg{\tilde z}=[11]$, we have the vanishing anticommutators
\bea\label{ccr1extra2}
&\{{\tilde z},{\tilde\psi}\}=\{{\tilde z},{\tilde\xi}\}=\{p_{\tilde z},{\tilde\psi}\}=\{p_{\tilde z},{\tilde\xi}\}=0.&
\eea
One can therefore set 
\bea\label{ccr1extra3}
{\tilde z} &=& y{\tilde \rho},
\eea
where $y$ is a standard coordinate and ${\tilde \rho}$ is a matrix anticommuting with ${\tilde \psi},{\tilde\xi}$.

\par
In analogy with the $(1,2,1)_{[00]}$ model, two sets of triples of hermitian $4\times 4$ matrices can be defined.
They can be expressed, in terms of the matrices defined in (\ref{splitquat}),  as
\bea\label{rep11a}
&{\tilde \psi}_A= \frac{1}{\sqrt 2} Y\otimes I,\quad {\tilde \xi}_A= \frac{i}{\sqrt 2} Y\otimes A,\quad {\tilde \rho}_A= i X\otimes A &
\eea
and
\bea\label{rep11b}
&{\tilde \psi}_B= \frac{1}{\sqrt 2} Y\otimes X,\quad {\tilde \xi}_B= \frac{i}{\sqrt 2} A\otimes Y,\quad {\tilde \rho}_B= -i I\otimes A.&
\eea
The hermitian supercharges ${\widetilde Q}_{10}^{A,B}$, ${\widetilde Q}_{01}^{A,B}$ are obtained as the quantized version of the
Noether charges ${\widecheck Q}_{10}$, ${\widecheck Q}_{01}$ given in (\ref{NchargeHamilton}) by using the representations (\ref{rep11a},\ref{rep11b}), respectively.
We have
\bea
{\widetilde Q}_{10}^A&=&{\tilde\rho}_A{\tilde\xi}_A\cdot \partial_y+{\tilde\psi}_A\cdot {\overline W}(y),\nn\\
{\widetilde Q}_{10}^B&=&{\tilde\rho}_B{\tilde\xi}_B\cdot \partial_y+{\tilde\psi}_B\cdot {\overline W}(y),\nn\\
{\widetilde Q}_{01}^A&=&{\tilde\rho}_A{\tilde\psi}_A\cdot \partial_y+{\tilde\xi}_A\cdot {\overline W}(y),\nn\\
{\widetilde Q}_{01}^B&=&{\tilde\rho}_B{\tilde\psi}_B\cdot \partial_y+{\tilde\xi}_B\cdot {\overline W}(y).\nn\\
\eea
The supercharges  satisfy 
\bea
&
({\widetilde Q}_{10}^A)^\dagger={\widetilde Q}_{10}^A,\quad 
({\widetilde Q}_{10}^B)^\dagger={\widetilde Q}_{10}^B,\quad 
({\widetilde Q}_{01}^A)^\dagger={\widetilde Q}_{01}^A,\quad 
({\widetilde Q}_{01}^B)^\dagger={\widetilde Q}_{01}^B.&
\eea
The four supercharges are square roots of the hermitian quantum Hamiltonian ${\widetilde H}$, recovered from the anticommutators
\bea
&\{{\widetilde Q}_{10}^A,{\widetilde Q}_{10}^A\}= \{{\widetilde Q}_{10}^B,{\widetilde Q}_{10}^B\}= \{{\widetilde Q}_{01}^A,{\widetilde Q}_{01}^A\}= \{{\widetilde Q}_{01}^B,{\widetilde Q}_{01}^B\}= 2{\widetilde H},\qquad {\widetilde H}^\dagger ={\widetilde H}.&
\eea
The quantum Hamiltonian ${\widetilde H}$ coincides with the quantum Hamiltonian $H$ given in (\ref{qham12100})
when setting ${\overline W}(y)=W(y)$. This implies that the ${\mathbb Z}_2\times {\mathbb Z}_2$-graded models $(1,2,1)_{[00]}$ and $(1,2,1)_{[11]}$, despite being classically different, produce after canonical quantization the same quantum theory.
{\color{black}
This is explained by the algebraic structure of the quantized operators. 
The representation of the relations (\ref{CCR1case11}, \ref{ccr1extra1}, \ref{ccr1extra2}, \ref{ccr1extra3}) is equivalent to the representation of the relations (\ref{CCR1}, \ref{CCR2}, \ref{anticommutatorrelations}). }

\section{Symmetries of the single-multiplet quantum Hamiltonian}

We present here the different graded symmetries of the quantum Hamiltonian $H$ given in (\ref{qham12100}). Since it coincides with the Hamiltonian derived from the quantization of the $(1,2,1)_{[11]}$ model, it is sufficient to discuss
the symmetry operators obtained from the quantum $(1,2,1)_{[00]}$ theory. We recall that, while  the Hamiltonian does not
depend on the chosen triple of hermitian matrices (\ref{triplesherm}), the Noether supercharges introduced in 
(\ref{fournoether}), on the other hand, depend on the given choice. By construction, each one of the four operators
$Q_{10}^A, Q_{10}^B, Q_{01}^A, Q_{01}^B$ in formulas  (\ref{fournoether}) and (\ref{absupercharges}) is a conserved symmetry operator. Therefore it makes sense to introduce,  for each such pair of operators in the ${\cal G}_{10}$ and ${\cal G}_{01}$ sectors, the induced ${\mathbb Z}_2\times{\mathbb Z}_2$-graded superalgebra. \par The results are the following:\par
~\par
{\it i}) {\it two copies of the Beckers-Debergh algebra.} \par 
~\par
By taking a pair of operators defined by the same triple of hermitian matrices, either ``$A$" or ``$B$",  
 we obtain two separate realizations of the Beckers-Debergh algebra, namely the ${\mathbb Z}_2\times {\mathbb Z}_2$-graded superalgebra
with vanishing ${\cal G}_{11}$ sector, see (\ref{z2z2becdeb}).\par The two copies of Beckers-Debergh algebras are respectively given by the two sets of three operators, ${\frak S}_A$ and ${\frak S}_B$,
\bea
{\frak S}_A = \{ Q_{10}^A, Q_{01}^A, H\} &{\textrm{and}}&
{\frak S}_B= \{Q_{10}^B, Q_{01}^B, H\},
\eea
which contain the common Hamiltonian $H$, with $H\in {\cal G}_{00}$. \par
In both cases the (\ref{z2z2becdeb}) (anti)commutators are satisfied since, besides the anticommutators in (\ref{squarerootsof}),  namely
$\{Q_{10}^A,Q_{10}^A\}= \{Q_{10}^B,Q_{10}^B\}= 2H$, all commutators
are vanishing:
\bea
[H, Q_{10}^A]=[H,Q_{01}^A]=[Q_{10}^A,Q_{01}^A]=0&{\textrm{and}}& 
[H, Q_{10}^B]=[H,Q_{01}^B]=[Q_{10}^B,Q_{01}^B]=0;
\eea
~\par
{\it ii}) {\it two copies of the one-dimensional ${\mathbb Z}_2\times {\mathbb Z}_2$-graded supertranslation algebra.} \par 
~\par
By taking a``mixed" pair of operators constructed from the two different triples (``$A$" and ``$B$") of the hermitian matrices (\ref{triplesherm}),  
 we obtain two, conveniently normalized, separate realizations of the one-dimensional ${\mathbb Z}_2\times {\mathbb Z}_2$-graded supertranslation algebra (\ref{z2z2super}). In both cases the ${\cal G}_{11}$ sector is nonvanishing.\par 
The two copies of the supertranslation algebra are respectively spanned by the two sets of four operators ${\frak S}_1$ and ${\frak S}_2$, given by
\bea
{\frak S}_1 = \{ Q_{10}^B, Q_{01}^A, H, Z\} &{\textrm{and}}&
{\frak S}_2= \{Q_{10}^A, Q_{01}^B, H,{\overline Z} \}.
\eea
Their respective  ${\cal G}_{11}$ sectors are spanned by the hermitian operators $Z$ and ${\overline Z}$,
\bea
Z=Z^\dagger, && {\overline Z}={\overline Z}^\dagger.
\eea
We have
\bea
{Z}&=& \left(\begin{array}{cccc}0&-\partial_y^2+W^2+W'&0&0\\ -\partial_y^2+W^2+W'&0&0&0\\0& 0&0&-\partial_y^2+W^2-W'\\ 0&0&-\partial_y^2+W^2-W'&0\end{array}\right)\nonumber\\&&
\eea
and
\bea
{\overline{Z}}&=&\left(\begin{array}{cccc}0&\partial_y^2-W^2-W'&0&0\\ \partial_y^2-W^2-W'&0&0&0\\ 0&0&0&-\partial_y^2+W^2-W'\\ 0&0&-\partial_y^2+W^2-W'&0\end{array}\right).\nonumber\\&&
\eea
The nonvanishing (anti)commutators of the  ${\mathbb Z}_2\times{\mathbb Z}_2$-graded superalgebra ${\frak S}_1$ are
\bea
& \{Q_{10}^B,Q_{10}^B\}= \{Q_{01}^A,Q_{01}^A\}=  2H, \qquad [Q_{10}^B,Q_{01}^A]=iZ.&
\eea
The nonvanishing (anti)commutators of the ${\mathbb Z}_2\times{\mathbb Z}_2$-graded superalgebra  ${\frak S}_2$ are
\bea
& \{Q_{10}^A,Q_{10}^A\}= \{Q_{01}^B,Q_{01}^B\}=  2H, \qquad [Q_{10}^A,Q_{01}^B]=i{\overline Z};&
\eea
\par
~\par
{\it iii}) {\it the ${\mathbb Z}_2$-graded superalgebra of the ${\cal N}=4$ supersymmetric quantum mechanics.} \par 
~\par
Besides the ${\mathbb Z}_2\times {\mathbb Z}_2$-graded symmetry algebras, the Hamiltonian $H$ possesses a ${\mathbb Z}_2$-graded symmetry, making it 
an example of ${\cal N}=4$ supersymmetric quantum mechanics \cite{wit} satisfying
\bea\label{n4super}
&\{Q_i,Q_j\}= 2\delta_{ij} H, \qquad [H, Q_i]=0,\qquad {\textrm{for}}\qquad i,j=1,2,3,4.&
\eea
The two sets of mixed Noether supercharges operators, $Q_{10}^B,Q_{01}^A$ and  $Q_{10}^A,Q_{01}^B$, define
two copies of superalgebras of the ${\cal N}=2$ supersymmetric quantum mechanics, since their respective anticommutators are both vanishing
\bea
\{Q_{10}^B,Q_{01}^A\}=0 \qquad &{\textrm{and}}&\qquad 
\{Q_{10}^A,Q_{01}^B\}=0.
\eea
In order to get the ${\cal N}=4$ superalgebra (\ref{n4super}), two extra supersymmetry operators, which do not coincide with the Noether supercharges (\ref{absupercharges}), have to be added. A convenient presentation of
the $4$ operators $Q_i$ satisfying (\ref{n4super}) is given by
\bea\label{n4operators}
Q_1&=&Q_{10}^B=-\frac{i}{\sqrt{2}}(Y\otimes X\cdot\partial_y+A\otimes X\cdot W(y)),\nn\\
Q_2&=&Q_{01}^A=\frac{1}{\sqrt{2}}(Y\otimes A\cdot\partial_y+A\otimes A\cdot W(y)),\nn\\
Q_3&=&-\frac{i}{\sqrt{2}}(Y\otimes Y\cdot\partial_y+A\otimes Y\cdot W(y)),\nn\\
Q_4&=&\frac{1}{\sqrt{2}}(A\otimes I\cdot\partial_y+Y\otimes I\cdot W(y)).\nn\\
\eea
The matrices $A,X,Y,I$ have been introduced in (\ref{splitquat}).

{\color{black}
The emergence of the ${\cal N}=4$ supersymmetry is a purely quantum phenomenon since the fields in the  $\mathbb{Z}_2\times\mathbb{Z}_2$-graded classical model have a different grading with respect to the supersymmetric fields. 
The classical ${\cal N}=4$ supersymmetric model for the Hamiltonian (\ref{qham12100}) is that of the $ (1,4,3)$ supermultiplet (see, e.g., the $D(2,1;\alpha)$ superconformal model discussed in\cite{CHT}). 
The coincidence of the two versions of the quantum models is understood by taking into account that both the classical ${\cal N}=4$ $(1,4,3)$ supermechanics and the classical $\mathbb{Z}_2\times\mathbb{Z}_2$-graded 
symmetry require, upon quantization, a $ 4 \times 4$ quantum Hamiltonian. }

\section{${\mathbb Z}_2\times{\mathbb Z}_2$-graded interacting multiparticle Hamiltonians}

 We now further apply our scheme to derive 
a new class of ${\mathbb Z}_2\times{\mathbb Z}_2$-graded quantum Hamiltonians. Specifically, we present the quantum systems obtained by quantizing the classical actions of several, interacting, $(1,2,1)_{[00]}$ multiplets. The construction of these classical models was presented in \cite{AKT1} (see subsection {\bf 5.3} of that paper). In particular,
the invariant action ${\cal S}_{2P}=\int dt {\cal L}_{2P}$ of the $2$-particle case is expressed, in real time $t$, in terms of the Lagrangian
\bea\label{twoplagr}
{\cal L}_{2P}&=& \frac{1}{2}[g_{11}({\dot x}_1^2-z_1^2+i\psi_1{\dot\psi}_1-i\xi_1{\dot\xi}_1)+g_{22}({\dot x}_2^2-z_2^2+i\psi_2{\dot\psi}_2-i\xi_2{\dot\xi}_2)+\nonumber\\
&&g_{12}(2{\dot x}_1{\dot x}_2-2z_1z_2+i\psi_1{\dot\psi}_2+i\psi_2{\dot\psi}_1-i\xi_1{\dot \xi}_2-i\xi_2{\dot \xi}_1)-g_{111}z_1\psi_1\xi_1-g_{222}z_2\psi_2\xi_2+\nonumber\\&&g_{112}(-z_2\psi_1\xi_1-z_1(\psi_1\xi_2+\psi_2\xi_1))+
g_{221}(-z_1\psi_2\xi_2-z_2(\psi_1\xi_2+\psi_2\xi_1))+\nonumber\\&&\lambda_1\mu z_1+\lambda_2\mu z_2].
\eea
The component fields of the two multiplets (respectively denoted as $x_1,\psi_1,\xi_1,z_1$ and $x_2,\psi_2,\xi_2,z_2$) transform independently under the ${\mathbb Z}_2\times{\mathbb Z}_2$-graded transformations (\ref{QQZtransf00}). The fields $x_1(t),x_2(t)$ describe the propagating bosons, while $z_1(t),z_2(t)$ describe the auxiliary bosons.  The Lagrangian depends on the prepotential function $g(x_1,x_2)$. The functions $g_{ij}(x_1,x_2)$ are  interpreted as the metric of the two-dimensional target manifold. The metric is constrained to satisfy the equation
\bea\label{metricprepot}
\qquad\qquad \qquad g_{ij}(x_1,x_2)& = &\partial_{x_i} \partial_{x_j} g(x_1,x_2),\qquad {\textrm{for}}\quad i,j=1,2,
\eea
in terms of the prepotential function $g(x_1,x_2)$. \par
The condition
\bea\label{interactingcond}
g_{12} (x_1,x_2) &\neq& 0
\eea
is necessary in order to have interacting multiplets. \par Finally, the linear terms in $z_1,z_2$ in the last line of the right hand side of (\ref{twoplagr}) depend on the $[11]$-graded coupling constant $\mu$, while $\lambda_1,\lambda_2\in {\mathbb R}$ are arbitrary real parameters.\par
The construction of invariant classical actions for  $n>2$ interacting $(1,2,1)_{[00]}$ multiplets is a straightforward extension of the $n=2$ procedure.

\subsection{Constant kinetic basis and classical Hamiltonian formulation}

The quantization of the (\ref{twoplagr}) action requires repeating the steps discussed in Section {\bf 3} and {\bf 4} concerning the quantization of the single multiplet Lagrangian (\ref{p00model}). Here, we limit ourselves to discuss the main relevant differences with respect to this case. \par

{\color{black}
The passage to the ``constant kinetic basis" is more involved for the $2$-particle case since the prepotential $g(x_1,x_2)$ depends on two coordinates. 
It is known in supermechanics that the passage to a constant kinetic basis is guaranteed if the prepotential depends on a single coordinate \cite{CHT}. 
The extension to $2$-particle $\mathbb{Z}_2 \times \mathbb{Z}_2$-graded mechanics case is in principle made possible by introducing two smooth and differentiable functions  $u(x_1,x_2)$
and $v(x_1,x_2)$; at least locally the transformations
}
\bea
x_1\mapsto u(x_1,x_2), &&x_2\mapsto v(x_1,x_2),
\eea
are assumed to be invertible. 
\par
The time derivatives are
\bea \label{timeder}
{\dot u} = u_1{\dot x}_1+u_2{\dot x}_2, && {\dot v} = v_1{\dot x}_1+v_2{\dot x}_2.
\eea
They are chosen so that the constant kinetic term
\bea\label{2pconstantkin}
K&=&\frac{1}{2}({\dot u}^2+{\dot v}^2)
\eea
reproduces the kinetic term for the (\ref{twoplagr}) propagating bosons, given by
\bea
K&=& \frac{1}{2}( g_{11}{\dot x}_1^2 +g_{22}{\dot x}_2^2+2g_{12}{\dot x}_1{\dot x}_2).
\eea
This is obtained with the identifications
\bea\label{identif}
g_{11} &=& u_1^2+v_1^2,\nonumber\\
g_{22} &=& u_2^2+v_2^2,\nonumber\\
g_{12}&=& u_1u_2+v_1v_2.
\eea
For the above $g_{ij}$ metric the Hessian $G=det(g_{ij})$ is
\bea\label{hessiang}
G&=& g_{11}g_{22}-g_{12}^2= (u_1v_2-u_2v_1)^2.
\eea
Being derived in terms of the prepotential $g(x_1,x_2)$ and satisfying the (\ref{metricprepot}) equation, the metric $g_{ij}$ satisfies the constraints
\bea
g_{11,2} &=&g_{12,1}, \nonumber\\ g_{22,1}&=& g_{12,2}.
\eea
Under the identifications (\ref{identif}) these constraints imply two nonlinear equations for the partial derivatives of $u$ and $v$. They are, respectively,
\bea\label{nonlinconstra}
C_1 &\equiv& u_1u_{12}-u_2u_{11}+v_1v_{12}-v_2v_{11}=0,\nn\\
C_2&\equiv& u_1u_{22}-u_2u_{12}+v_1v_{22}-v_2v_{12}=0.
\eea
It is worth mentioning that the (\ref{nonlinconstra}) constraints for $u,v$ admit nontrivial solutions. \par
As an example, the cubic polynomials
\bea
u(x_1,x_2)= x_1(1+\alpha x_1^2+3\alpha x_2^2), && v(x_1,x_2) = x_2(1+3\alpha x_1^2+\alpha x _2^2),
\eea
satisfy (\ref{nonlinconstra}) and induce, for any real $\alpha\in{\mathbb R}$, the metric $g_{ij}=\partial_i\partial_jg(x_1,x_2)$ obtained from the prepotential
\bea
g(x_1,x_2) &=& \frac{1}{2}(x_1^2+x_2^2)+\frac{\alpha}{2}(x_1^4+x_2^4)+\frac{3}{10}\alpha^2(x_1^6+x_2^6)+\nn\\&&
3\alpha x_1^2x_2^2+\frac{9}{2}\alpha^2x_1^2x_2^2(x_1^2+x_2^2).
\eea
The two-dimensional constant Euclidean metric is recovered at $\alpha=0$.\par
For the $[10]$-graded odd fields $\psi_1,\psi_2$ the change of variables
\bea\label{fermionicchange}
\left(\begin{array}{c} \psi_u\\ \psi_v\end{array}\right )&=&\left(\begin{array}{cc}u_1&u_2\\v_1&v_2\end{array}\right)
\left(\begin{array}{c} \psi_1\\ \psi_2\end{array}\right )
\eea
guarantees that the fermionic constant kinetic term $K_f$,
\bea
K_f &=& \frac{i}{2}(\psi_u{\dot\psi_u}+\psi_v{\dot\psi_v})=\nn\\&=&
\frac{i}{2}(g_{11}\psi_1{\dot\psi_1}+g_{22}\psi_2{\dot\psi_2}+g_{12}(\psi_1{\dot\psi_2}+\psi_2{\dot\psi_2})+(C_1{\dot x_1}+C_2{\dot x}_2)\psi_1\psi_2),
\eea
reproduces the fermionic kinetic term for $\psi_1,\psi_2$ in (\ref{twoplagr}) once provided that the nonlinear constraints
$C_1=C_2=0$ from formulas (\ref{nonlinconstra}) are satisfied.\par
An analogous change of variables is made by replacing in equation (\ref{fermionicchange}) the fields $\psi_1,\psi_2$ and $\psi_u,\psi_v$ with the $[01]$-graded fields $\xi_1,\xi_2$ and $\xi_u,\xi_v$, respectively. \par
From now on everything proceeds as in the single multiplet case. 
After solving the algebraic equations of motion for the auxiliary fields $z_1,z_2$, we then introduce canonical variables, Poisson brackets, Dirac brackets and the Hamiltonian formalism in terms of the new component fields
of the constant kinetic basis.\par
In particular in this basis the analogs of formulas (\ref{Ncharge00-4}) now read, for the two-particle  Noether charges 
${\widehat Q}_{2P;10}$, ${\widehat Q}_{2P;01}$, as
\bea\label{2pclassicsupercharges}
{\widehat Q}_{2P;10} &=& p_u\psi_u+p_v\psi_v +i\mu(W_u{\xi_u}+W_v\xi_v),
\nn\\
{\widehat Q}_{2P; 01} &=& p_u\xi_u+p_v\xi_v -i\mu(W_u{\psi_u}+W_v\psi_v),
\eea
where $p_u,p_v$ are the conjugate momenta $p_u\equiv{\dot u}$, $p_v\equiv {\dot v}$.\par
Instead of a single field $W$ as in (\ref{newnoetherwithW}), we have now two fields, $W_u$, $W_v$.  They are derived from
the $\lambda_1,\lambda_2$ terms in the Lagrangian (\ref{twoplagr}) and satisfy
\bea
W_u\psi_u+W_v\psi_v= \lambda_1\psi_1+\lambda_2\psi_2, && W_u\xi_u+W_v\psi_v\xi_v=\lambda_1\xi_1+\lambda_2\xi_2,
\eea
so that the analog of formula (\ref{Wterm}) is provided by
\bea\label{WuWv}
W_u = \frac{1}{\widetilde G}(\lambda_1v_2-\lambda_2v_1), &&
W_v = \frac{1}{\widetilde G}(-\lambda_1u_2+\lambda_2u_1).
\eea
One should note that the suffices $u,v$ denote the two different fields $W_u,W_v$ and are not a symbol of derivation.\par
In the above expressions
\bea
{\widetilde G}&=& u_1v_2-u_2v_1
\eea
is the determinant of the transformation matrix entering  the right hand side of (\ref{fermionicchange}). Due to (\ref{identif}), we have that the Hessian $G$ given in (\ref{hessiang}) is
\bea
G&=& {\widetilde G}^2.
\eea
In the $g_{12}=0$ non-interacting case the functions $u,v$ can be chosen as $u=u(x_1)$, $v=v(x_2)$. Therefore
$u_2=v_1=0$ and the expressions for $W_u$, $W_v$ are simpler. Under this assumption we get:
\bea\label{uvderivatives}
&u_2=v_1=0 \quad {\textrm{imply}}\quad W_u =\frac{\lambda_1}{u_1},\quad W_v=\frac{\lambda_2}{v_2}\quad {\textrm{and}}\quad \partial_uW_v=\partial_vW_u=0.&
\eea
The weaker condition $\partial_uW_v=\partial_vW_u$ for the interacting case is a result, as discussed below in subsection {\bf 6.3},
of the quantization procedure requiring a matrix representation of the variables $\psi_u,\psi_v,\xi_u,\xi_v$.
\par
The non-vanishing Dirac brackets of the conjugate variables are
\bea
&\{u,p_u\}_D=\{v,p_v\}_D=1,\quad \{\psi_u,\psi_u\}_D=\{\psi_v,\psi_v\}_D=-i,\quad \{\xi_u,\xi_u\}_D=\{\xi_v,\xi_v\}_D=i.&
\eea
The classical Hamiltonian ${\widehat H}_{2P}$ can be read, in analogy with formula (\ref{DBclassicalalg}), through the Dirac brackets
\bea
&\{{\widehat Q}_{2P;01}, {\widehat Q}_{2P;01}\}_D=-\{{\widehat Q}_{2P;10}, {\widehat Q}_{2P;10}\} = 2i{\widehat H}_{2P}.&
\eea
The extension of this construction to the case of $n>2$ interacting multiplets $(1,2,1)_{[00]}$ is straightforward. One obtains $n$ pairs of conjugated variables for the propagating bosons, $n$ $[10]$-graded fields $\psi_i$ and $n$ $[01]$-graded fields $\xi_i$.

\subsection{Matrix representations of the quantum (anti)commutators}

The construction of the quantum theory requires matrices which solve the (anti)commutators which extend the set of single-particle relations (\ref{anticommutatorrelations}) induced by the Dirac brackets. In the passage from a single to $n$ multiplets we have $n$ matrices denoted as ${\overline\psi}_i$, $n$ matrices denoted as ${\overline\xi}_i$, with $ {\overline \psi}_i\in {\cal G}_{10}$, ${\overline \xi}_i\in {\cal G}_{01}$ ($i=1,2,\ldots,n)$ and the single matrix $\mu\in{\cal G}_{11}$. They satisfy the (anti)commutators
\bea\label{nanticommutatorrelations}
&\{{\overline \psi}_i,{\overline\psi}_j\}=-\{{\overline\xi}_i,{\overline\xi}_j\}=\delta_{ij}\cdot{\mathbb I},\quad [{\overline \psi}_i,{\overline \xi}_j]=0,\quad \{{\overline \psi}_i,\mu\}=\{{\overline \xi}_i,\mu\}=0\qquad (i,j=1,\ldots n),&\nn\\&&
\eea
where ${\mathbb I}$ is the identity matrix of proper size. 
This system is minimally solved by $2^{n+1}\times 2^{n+1}$ real matrices. 
The matrix sectors ${\cal G}_{10}, {\cal G}_{01}, {\cal G}_{11}$ can be read from formula (\ref{gradedmatrices}) with the entries given by $2^{n-1}\times 2^{n-1}$ blocks. We present the $n=2,3$ solutions.
\par
We have at first to take into account the gradings $ {\overline \psi}_i\in {\cal G}_{10}$, ${\overline \xi}_i\in {\cal G}_{01}$, $\mu\in{\cal G}_{11}$ and the (anti)symmetry of these real matrices. Therefore, up to a normalization factor, for $n=2$ the matrices should be respectively picked up from the sets
\bea\label{2psolutiona}
{\overline\psi}_i &:&
Y\otimes I\otimes I, ~Y\otimes X\otimes I, ~Y\otimes I\otimes X, ~Y\otimes X\otimes X, \nn\\&& A\otimes I\otimes A, ~ A\otimes X\otimes A, ~Y\otimes I\otimes Y, ~Y\otimes X\otimes Y;\nonumber\\
{\overline\xi}_i&:& Y\otimes A\otimes I,~ Y\otimes A\otimes X,~ Y\otimes A\otimes Y,~ A\otimes A\otimes A,\nn\\&& A\otimes Y\otimes I, ~A\otimes Y\otimes X, ~A\otimes Y\otimes Y, ~Y\otimes Y\otimes A;\nonumber\\
\mu&:& X\otimes A\otimes X, ~X\otimes A\otimes Y, ~X\otimes A\otimes I, ~X\otimes Y\otimes A,
\nn\\&& I\otimes A\otimes X,~ I\otimes A\otimes Y,~ I\otimes A\otimes I,~ I\otimes Y\otimes A,
\eea 
where  $A,X,Y,I$ are the $2\times 2$ matrices given in (\ref{splitquat}).\par
Consistent $n=2$ solutions of the (anti)commutator relations (\ref{nanticommutatorrelations}) are found by setting,
e.g.:
\bea\label{2psolutionb}
&{\overline\psi}_1=\frac{1}{\sqrt{2}}Y\otimes X\otimes X,\quad
{\overline\psi}_2=\frac{1}{\sqrt{2}} Y\otimes X\otimes Y,
\quad {\textrm{then}}\quad  
{\overline \xi}_1=\frac{1}{\sqrt{2}}A\otimes Y\otimes I,\quad
{\overline \xi}_2=\frac{1}{\sqrt{2}} Y\otimes Y\otimes A,&\nn\\&
  {\textrm{while}} \quad \mu \quad {\textrm{is either}} \quad \mu= X\otimes Y\otimes A\quad {\textrm{or}}\quad \mu=I\otimes A\otimes I&
\eea
and, alternatively,
\bea\label{2psolutionc}
&{\overline\psi}_1=\frac{1}{\sqrt{2}}Y\otimes I\otimes X,\quad
{\overline\psi}_2=\frac{1}{\sqrt{2}} Y\otimes I\otimes Y,
\quad {\textrm{then}}\quad  
{\overline \xi}_1=\frac{1}{\sqrt{2}}Y\otimes A\otimes I,\quad
{\overline \xi}_2=\frac{1}{\sqrt{2}} A\otimes A\otimes A,&\nn\\&
  {\textrm{while}} \quad \mu \quad {\textrm{is either}} \quad \mu= X\otimes A\otimes I\quad {\textrm{or}}\quad \mu=I\otimes Y\otimes A.&
\eea
For the $3$-particle case, an $n=3$ solution of the (anti)commutators (\ref{nanticommutatorrelations}) is given by
\bea\label{n3solutions}
&{\overline\psi}_1=\frac{1}{\sqrt{2}}\cdot Y\otimes I\otimes X\otimes X ,\quad
{\overline\psi}_2=\frac{1}{\sqrt{2}}\cdot Y\otimes I\otimes X\otimes Y,\quad{\overline\psi}_3=\frac{1}{\sqrt{2}}\cdot Y\otimes I\otimes Y\otimes I,&\nn\\
&{\overline\xi}_1=\frac{1}{\sqrt{2}}\cdot A\otimes A\otimes A\otimes I,\quad{\overline\xi}_2=\frac{1}{\sqrt{2}}\cdot Y\otimes Y\otimes Y\otimes A,\quad{\overline\xi}_3=\frac{1}{\sqrt{2}}\cdot Y\otimes A\otimes I\otimes I,&\nn\\
&{\mu}= I\otimes Y\otimes A\otimes I.&
\eea
We are now in the position to present the ${\mathbb Z}_2\times {\mathbb Z}_2$-graded hermitian quantum Hamiltonians for $n=2,3$ (both interacting and non-interacting) $(1,2,1)_{[00]}$ multiplets.  The general construction
for $n\geq 4$ follows the scheme here outlined.

\subsection{${\mathbb Z}_2\times{\mathbb Z}_2$-graded quantum Hamiltonians of two interacting particles}  

It is convenient to indicate here as $x,y$ the coordinates associated with the propagating bosons that,
in the constant kinetic basis, were previously denoted as $u,v$.  \par
The hermitian, constant matrices  derived from the (\ref{2psolutionb}) solutions can be expressed as
\bea\label{2parta}
&\psi_1^A =\frac{1}{\sqrt 2}\cdot Y\otimes X\otimes X, \quad \psi_2^A=\frac{1}{\sqrt 2}\cdot Y\otimes X\otimes Y,\qquad\qquad\qquad\qquad\qquad~~ &\nn\\
&\xi_1^A=\frac{i}{\sqrt 2}\cdot A\otimes Y\otimes I,\quad \xi_2^A=\frac{i}{\sqrt 2} \cdot Y\otimes Y\otimes A,\qquad \mu^A=-i \cdot I\otimes A\otimes I.&
\eea
The hermitian, constant matrices  derived from the (\ref{2psolutionc}) solutions can be expressed as
\bea\label{2partb}
&\psi_1^B =\frac{1}{\sqrt 2}\cdot Y\otimes I\otimes X, \quad \psi_2^B=\frac{1}{\sqrt 2}\cdot Y\otimes I\otimes Y,\qquad\qquad\qquad\qquad\qquad~~ &\nn\\
&\xi_1^B=\frac{i}{\sqrt 2}\cdot Y\otimes A\otimes I,\quad \xi_2^B=\frac{i}{\sqrt 2} \cdot A\otimes A\otimes A,\qquad \mu^B=i \cdot X\otimes A\otimes I.&
\eea

The two-particle hermitian Noether supercharges derived from (\ref{2parta}) and (\ref{2partb}) and corresponding to the quantization of formulas (\ref{2pclassicsupercharges}) are
\bea\label{2psupercharges}
Q_{2P;10}^{A} &=& -i(\psi_1^A\partial_x+\psi_2^A\partial_y+W_1\mu^A\xi_1^A+W_2\mu^A\xi_2^A),\nn\\
Q_{2P;01}^{A} &=& -i(\xi_1^A\partial_x+\xi_2^A\partial_y+W_1\mu^A\psi_1^A+W_2\mu^A\psi_2^A),\nn\\
Q_{2P;10}^{B} &=& -i(\psi_1^B\partial_x+\psi_2^B\partial_y+W_1\mu^B\xi_1^B+W_2\mu^B\xi_2^B),\nn\\
Q_{2P;01}^{B} &=&-i(\xi_1^B\partial_x+\xi_2^B\partial_y+W_1\mu^B\psi_1^A+W_2\mu^B\psi_2^B) .
\eea

They depend on the real functions $W_1(x,y)$ and $W_2(x,y)$. In the non-interacting case we have
 $\partial_y W_1= \partial_x W_2=0$, so that $W_1\equiv W_1(x)$, $W_2\equiv W_2(y)$. The selection of the normalized matrices $\mu^A, \mu^B$ in (\ref{2psupercharges}) and not of their alternative choices respectively presented  in (\ref{2psolutionb}) and (\ref{2psolutionc}) is made to ensure that the non-interacting Hamiltonian is a diagonal operator.\par
 In the interacting case the weaker condition
\bea\label{2pintercond}
\partial_y W_1(x,y) &=& \partial_x W_2(x,y),
\eea
which is solved by the positions
\bea\label{2pintercondbis}
W_1 =\partial_x f(x,y)=f_x, &\quad&W_2 =\partial_y f(x,y)=f_y
\eea
in terms of the unconstrained function $f(x,y)$, has to be enforced.\par
The condition (\ref{2pintercond}) is derived, at the quantum level, by the requirement that the operators  $Q_{2P;10}^{A}, Q_{2P;10}^{B}, Q_{2P;01}^{A} , Q_{2P;01}^{B} $ are square roots of the same Hamiltonian $H_{2P}$. Since, e.g., in the matrix representation (\ref{2psolutionb}) we have
\bea\label{psixicond}
\psi_1^A\xi_2^A &\neq &\psi_2^A\xi_1^A,
\eea
the condition
\bea
(Q_{2P;10}^{A} )^2-(Q_{2P;01}^{A})^2&=&0
\eea
implies that the first order derivative  contributions, appearing on the left hand side, 

\bea
i\mu^A((\partial_xW_2-\partial_yW_1)\psi_1^A\xi_2^A-(\partial_xW_2-\partial_yW_1)\psi_2^A\xi_1^A)&=&0,
\eea 
should separately vanish for $\psi_1^A\xi_2^A$ and $\psi_2^A\xi_1^A$, thus leading to (\ref{2pintercond}). The condition (\ref{psixicond}) is representation-dependent and not necessarily implied by the classical derivation. \par
It should be stressed that, for a given
$f(x,y)$, the consistency of the ${\mathbb Z}_2\times {\mathbb Z}_2$-graded quantum theory does not require to solve the inverse equations which are induced, see (\ref{WuWv}), by the classical theory.\par
When (\ref{2pintercondbis}) is enforced
the  four operators given in (\ref{2psupercharges}) are square-roots of the two-particle Hamiltonian $H_{2P}$. They satisfy
\bea
&\{Q_{2P;10}^{A},Q_{2P;10}^{A}\}=\{Q_{2P;01}^{A},Q_{2P;01}^{A}\}=\{Q_{2P;10}^{B},Q_{2P;10}^{B}\}=\{Q_{2P;01}^{B},
Q_{2P;01}^{B}\}=2H_{2P},&
\eea
where
\bea\label{ham2p}
H_{2P}&=&{\footnotesize{\left(\begin{array}{cccccccc}H_0+V_{++}&0&0&0&0&0&0&0\\0&H_0+V_{--}&0&0&0&0&0&0\\
0&0&H_0+V_{++}&0&0&0&0&0\\0&0&0&H_0+V_{--}&0&0&0&0\\
0&0&0&0&H_0+V_{-+}&-f_{xy}&0&0\\0&0&0&0&-f_{xy}&H_0+V_{+-}&0&0\\
0&0&0&0&0&0&H_0+V_{-+}&-f_{xy}\\0&0&0&0&0&0&-f_{xy}&H_0+V_{+-}
\end{array} \right)}}\nonumber\\&&
\nn\\
{\textrm{with}}&&H_0= \frac{1}{2}(-\partial_x^2-\partial_y^2+f_x^2+f_y^2) \qquad {\textrm{and}}\qquad V_{\epsilon\delta}= \frac{1}{2}(\epsilon f_{xx}+\delta f_{yy}), \quad {\textrm{for}}\quad \epsilon,\delta=\pm 1.
\eea
By construction, the $2$-particle Hamiltonian (\ref{ham2p}) is hermitian. In the
\bea
f_{xy}\equiv \frac{\partial^2}{\partial_x\partial_y}f(x,y) &=&0
\eea
non-interacting case the Hamiltonian $H_{2P}$ is a diagonal operator.
\par
We can repeat for the $H_{2P}\in{\cal G}_{00}$ Hamiltonian the analysis of the ${\mathbb Z}_2\times{\mathbb Z}_2$-graded symmetries given in Section {\bf 5} for the single-particle Hamiltonian.\par
Two copies of the Beckers-Debergh algebra (\ref{z2z2becdeb}) are obtained from the two sets of three operators
\bea
{\frak S}_{2P,A} = \{ Q_{2P;10}^{A}, Q_{2P;01}^{A}, H_{2P}\} &{\textrm{and}}&
{\frak S}_{2P,B}= \{Q_{2P;10}^{B}, Q_{2P;01}^{B}, H_{2P}\}.
\eea
We have indeed vanishing commutators
\bea
[Q_{2P;10}^{A}, Q_{2P;01}^{A}]=0 \quad &{\textrm{and}}&\quad 
[Q_{2P;10}^{B}, Q_{2P;01}^{B}]=0.
\eea
The two sets of four hermitian  operators
\bea
{\frak S}_{2P,1}= \{ Q_{2P;10}^{B}, Q_{2P;01}^{A}, H_{2P}, Z_{2P}\} &{\textrm{and}}&
{\frak S}_{2P,2}= \{Q_{2P;10}^{A}, Q_{2P;01}^{B}, H_{2P},{\overline Z}_{2P} \}
\eea
produce two copies of the 
one-dimensional ${\mathbb Z}_2\times {\mathbb Z}_2$-graded supertranslation algebra (\ref{z2z2super}),
with nonvanishing (anti)commutators 
\bea\label{z2p}
& \{Q_{2P;10}^{B},Q_{2P;10}^{B}\}= \{Q_{2P;01}^{A},Q_{2P;01}^{A}\}=  2H_{2P}, \qquad [Q_{2P;10}^{B},Q_{2P;01}^{A}]=iZ_{2P}&
\eea
and, respectively,
\bea\label{z2poverlined}
& \{Q_{2P;10}^{A},Q_{2P;10}^{A}\}= \{Q_{2P;01}^{B},Q_{2P;01}^{B}\}=  2H_{2P}, \qquad [Q_{2P;10}^{A},Q_{2P;01}^{B}]=i{\overline Z}_{2P}.&
\eea
The explicit expression of the $Z_{2P}, {\overline Z}_{2P}\in {\cal G}_{11}$ operators can be read from the commutators in formulas 
(\ref{z2p}) and (\ref{z2poverlined}). We present, for completeness, $Z_{2P}$. It is given in terms of
the
$8\times 8$ matrices $E_{i,j}$ with entry $1$ at the intersection of the $i$-th row and $j$-th column and $0$ otherwise. We have
\bea
Z_{2P}&=& -(\partial_x^2+\partial_y^2-f_x^2-f_y^2)(-E_{1,3}+E_{2,4}-E_{3,1}+E_{4,2})+\nn\\
&& (\partial_x^2-\partial_y^2-f_x^2+f_y^2)(-E_{5,7}+E_{6,8}-E_{7,5}+E_{8,6})+\nn\\
&& -2(\partial_x\partial_y-f_xf_y)(E_{5,8}+E_{6,7}+E_{7,6}+E_{8,5})+\nn\\
&&2(f_x\partial_y-f_y\partial_x)(E_{5,8}-E_{6,7}+E_{7,6}-E_{8,5})+\nn\\
&&-(f_{xx}+f_{yy})(E_{1,3}+E_{2,4}+E_{3,1}+E_{4,2}+E_{5,7}+E_{6,8}+E_{7,5}+E_{8,6}).
\eea

\subsection{${\mathbb Z}_2\times{\mathbb Z}_2$-graded quantum Hamiltonians of three interacting particles}  

The $n=3$ solutions (\ref{n3solutions}) of the (anti)commutator relations (\ref{nanticommutatorrelations}) allow to define the ${\mathbb Z}_2\times {\mathbb Z}_2$-graded, three-particle, interacting quantum Hamiltonian $H_{3P}$.\par
The three coordinates are now labeled, for simplicity, as $x,y,z$.  The hermitian, constant matrices  expressed in terms of the
${\overline\psi}_1, {\overline\psi}_2, {\overline\psi}_3, {\overline\xi}_1, {\overline\xi}_2, {\overline\xi}_3, \mu $ matrices defined in (\ref{n3solutions}) are
\bea
&\psi_1 ={\overline\psi}_1, \quad \psi_2={\overline\psi}_2, \quad  \psi_3={\overline\psi_3},\quad
\xi_1 =i{\overline\xi}_1, \quad \xi_2=i{\overline\xi}_2, \quad  \xi_3=i{\overline\xi_3},\quad{\overline\mu}=i\mu.&
\eea
The hermitian supercharges, respectively  belonging to the ${\cal G}_{10}$ and ${\cal G}_{01}$ sectors,  are given by
\bea\label{3psupercharges}
Q_{3P;10}&=& -i(\psi_1\partial_x+\psi_2\partial_y+\psi_3\partial_z+f_x{\overline \mu}\xi_1+f_y{\overline \mu}\xi_2+f_z{\overline\mu}\xi_3),\nn\\
Q_{3P;01} &=& 
-i(\xi_1\partial_x+\xi_2\partial_y+\xi_3\partial_z+f_x{\overline \mu}\psi_1+f_y{\overline \mu}\psi_2+f_z{\overline\mu}\psi_3),
\eea
where $f(x,y,z)$ is an arbitrary function of the three coordinates.\par
They are the square roots of the hermitian three-particle Hamiltonian $H_{3P}$, given by
\bea\label{ham3p}
& \{Q_{3P;10},Q_{3P;10}\}= \{Q_{3P;01},Q_{3P;01}\}=  2H_{3P}.&
\eea
The Hamiltonian $H_{3P}$ is a $16\times 16$ matrix differential operator. It is the sum of a diagonal part $H_{diag}$ and of the off-diagonal terms $H_{off}$. In its turn the diagonal part is the sum of two terms, $H_0$ and $V$. We can set
\bea\label{ham3pint}
H_{3P} &=& H_{diag}+H_{off}, \qquad H_{diag}=H_0+V,
\eea
where
\bea
H_0&=& \frac{1}{2}(-\partial_{xx}^2-\partial_{yy}^2-\partial_{zz}^2+f_x^2+f_y^2+f_z^2)\cdot {\mathbb I}_{16},
\eea
while $V$ and $H_{off}$ are conveniently expressed in terms of the $16\times 16$ matrices $E_{i,j}$ with entry $1$ at the intersection of the $i$-th row and $j$-th column and $0$ otherwise. We get
\bea
V&=& V_{--+} (E_{1,1}+E_{7,7})+V_{+++} (E_{2,2}+E_{8,8})+V_{+--} (E_{3,3}+E_{5,5})+\nn\\&& V_{-+-} (E_{4,4}+E_{6,6})+
V_{+-+} (E_{9,9}+E_{15,15})+V_{-++} (E_{10,10}+E_{16,16})+\nn\\&&V_{---} (E_{11,11}+E_{13,13})+V_{++-} (E_{12,12}+E_{14,14}),
\eea
where
\bea
V_{\epsilon\delta\rho}&=& \frac{1}{2}(\epsilon f_{xx}+\delta f_{yy}+\rho f_{zz}), \qquad {\textrm{for}}\quad \epsilon,\delta,\rho=\pm 1.
\eea
The off-diagonal terms are
\bea\label{hoff3p}
H_{off}&=& f_{xy}(E_{3,4}+E_{4,3}+E_{5,6}+E_{6,5}+E_{9,10}+E_{10,9}+E_{15,16}+E_{16,15})+\nonumber\\
&&f_{xz}(-E_{1,3}-E_{3,1}+E_{5,7}+E_{7,5}+E_{10,12}+E_{12,10}-E_{14,16}-E_{16,14})+\nn\\&&
f_{yz}(-E_{1,4}-E_{4,1}+E_{6,7}+E_{7,6}-E_{9,12}-E_{12,9}+E_{14,15}+E_{15,14}).
\eea
The off-diagonal part of the Hamiltonian vanishes ($H_{off}=0$) if the three particles are not interacting, namely for
\bea
&f_{xy}=f_{xz}=f_{yz}=0.&
\eea
The set of three operators $Q_{3P;10}$, $Q_{3P;01}$, $H_{3P}$ close the Beckers-Debergh algebra (\ref{z2z2becdeb}) since the commutator between $Q_{3P;10}$, $Q_{3P;01}$ is vanishing:
\bea
[Q_{3P;10},Q_{3P;01}]&=&0.
\eea
\section{Conclusions}

In this paper we estalished the Hamiltonian formalism for classical ${\mathbb Z}_2\times{\mathbb Z}_2$-graded invariant
mechanical theories and performed their canonical quantization. We had to carefully specify how the necessary ingredients (Poisson and Dirac brackets, canonical variables, etc.) apply to graded fields. We worked out the cases of $(1,2,1)_{[00]}$ and $(1,2,1)_{[11]}$ ${\mathbb Z}_2\times{\mathbb Z}_2$-graded multiplets of component fields. Both these multiplets contain, see \cite{AKT1}, two types of fermionic fields, one ordinary boson
and one exotic boson (in the $(1,2,1)_{[00]}$ multiplet the ordinary boson is propagating while the exotic boson is an auxiliary field, the situation being reversed for the $(1,2,1)_{[11]}$ multiplet). The theories derived by these multiplets are the simplest ones. For the moment we left aside the quantization of the theories based on the other type
of ${\mathbb Z}_2\times{\mathbb Z}_2$-graded multiplet (denoted as ``$(2,2,0)$"
since it possesses two propagating bosons and two propagating fermions) introduced in \cite{AKT1}. Based on the quantization of its counterpart in ordinary supersymmetric quantum mechanics, see \cite{CHT}, we are expecting the $(2,2,0)$ multiplet to produce more complicated quantum Hamiltonians than the ones here derived. They should present, in particular, a spin-orbit interaction. This class of theories will be left for future investigations.\par
Concerning the two types of ${\mathbb Z}_2\times {\mathbb Z}_2$-graded $(1,2,1)$ multiplets we proved that the
quantization of the single $(1,2,1)_{[11]}$ multiplet produces the same $4\times 4$ matrix differential Hamiltonian induced by the quantization of $(1,2,1)_{[00]}$.  We point out that we do not have a general argument
that this should be necessarily the case for several interacting multiplets. The identification of the two Hamiltonians
for the two single-multiplet cases can be the byproduct of their simplicity. Loosely speaking the ${\mathbb Z}_2\times{\mathbb Z}_2$-graded symmetry, in this simple setting, does not leave room for an alternative, ${\mathbb Z}_2\times {\mathbb Z}_2$-graded, invariant Hamiltonian.\par 
The ${\mathbb Z}_2\times {\mathbb Z}_2$-graded quantum Hamiltonian introduced in \cite{BruDup} is recovered from the quantization of the classical invariant actions for a single multiplet (either $(1,2,1)_{[00]}$ or $(1,2,1)_{[11]}$)
which were presented in \cite{AKT1}. It turns out that this Hamiltonian is the $n=1$ representative of the class of ${\mathbb Z}_2\times {\mathbb Z}_2$-graded quantum Hamiltonians obtained by quantizing $n$ interacting multiplets, see formula (\ref{ham2p}) for $n=2$ and formulas (\ref{ham3pint}-{\ref{hoff3p}) for $n=3$.\par
The construction of multiparticle quantum Hamiltonians is particularly relevant because it helps answering the long-standing puzzle of the physical significance of a ${\mathbb Z}_2\times{\mathbb Z}_2$-graded symmetry. The single-particle quantum Hamiltonian (\ref{qham12100}) possesses different types of graded symmetries as discussed in Section {\bf 5}. Besides admitting ${\mathbb Z}_2\times{\mathbb Z}_2$-graded invariance, it is also an example of an ordinary ${\cal N}=4$ supersymmetric quantum mechanics, see (\ref{n4operators}). Its
${\mathbb Z}_2\times{\mathbb Z}_2$-graded symmetry is emergent, but the construction of the Hilbert space and the computation of the energy eigenvalues of the model do not require it. A radical new feature appears when considering multiparticle quantum Hamiltonians based on $n>1$ multiplets. In that case the ${\mathbb Z}_2\times{\mathbb Z}_2$-graded symmetry directly affects the statistics of the particles and the construction of the,
properly (anti)symmetrized, multiparticle wave functions. It implies measurable physical consequences (about the  energy eigenvalues and their degeneracies, the partition function and the chemical potentials, etc.). In \cite{top} the
experimentally testable consequences of the ${\mathbb Z}_2\times{\mathbb Z}_2$-graded parastatistics for
multiparticle quantum Hamiltonians have been presented.\par
{\color{black}
Indeed, it was shown in \cite{top}, by analyzing a simple oscillator model, that even if the multiparticle ${\mathbb Z}_2\times{\mathbb Z}_2$-graded and ${\cal N}=4$ supersymmetric quantum Hamiltonians coincide, the $\mathbb{Z}_2\times \mathbb{Z}_2$-graded and supersymmetric multiparticle Hilbert spaces differ. 
Furthermore, the measurement of some observables, applied on certain given states, allows to determine whether the multiparticle system under consideration is composed by
ordinary bosons/fermions (i.e., it is supersymmetric) or by parafermions (i.e., it is  $\mathbb{Z}_2\times \mathbb{Z}_2$-graded). This is made possible because the $\mathbb{Z}_2\times \mathbb{Z}_2$-graded parastatistics is encoded  in the braided tensor product introduced in \cite{maj}. \par
The results of the present work, together with those in \cite{AKT1}, have established the $ \mathbb{Z}_2\times \mathbb{Z}_2$-graded mechanics and its canonical quantization. This is a general framework that can now be applied in model
building and in testing the physical consequences of the $ \mathbb{Z}_2\times \mathbb{Z}_2$-graded invariance and its associated parastatistics. It allows a systematic investigation of various $\mathbb{Z}_2\times \mathbb{Z}_2$-graded systems. As examples we can mention the conformal mechanics, multiparticle classical and quantum systems with additional degrees of freedom (like spin)  and so on. 
}
\par

~\par
~\par
\par 
\newpage
{\Large{\bf Acknowledgments}}
{}~\par{}~\par

Z. K. and F. T. are grateful to the Osaka Prefecture University, where this work was completed, for hospitality.
 F. T. was supported by CNPq (PQ grant 308095/2017-0). 

~\par
~\par

  \renewcommand{\theequation}{A.\arabic{equation}}
  \setcounter{equation}{0}  

\textcolor{black}{
{\Large{\bf{Appendix: reminder of ${\mathbb Z}_2\times{\mathbb Z}_2$-graded superalgebras}}} }\par
~\par

We summarize for completeness, following\cite{AKT1}, the basic properties of  the ${\mathbb Z}_2\times{\mathbb Z}_2$-graded superalgebras and conventions used in the text.\par
A ${\mathbb Z}_2\times{\mathbb Z}_2$-graded Lie superalgebra ${\cal G}$ is decomposed as
\bea
{\cal G}&=& {\cal G}_{00}\oplus {\cal G}_{10}\oplus{\cal G}_{01}\oplus {\cal G}_{11}.
\eea
It is endowed with the operation $[\cdot,\cdot\}: {\cal G}\times {\cal G}\rightarrow {\cal G}$ satisfying, 
for any $g_a\in {\cal G}_{{\vec{\alpha}}}$,  the properties
\bea\label{z2z2brackets}
&[g_a,g_b\} = g_ag_b-(-1)^{{\vec{\alpha}}\cdot{\vec{\beta}}}g_bg_a,&\nonumber
\\
&(-1)^{{\vec{\gamma}}\cdot{\vec{\alpha}}}[g_a,[g_b,g_c\} \}+
(-1)^{{\vec{\alpha}}\cdot{\vec{\beta}}}[g_b,[g_c,g_a\} \}+
(-1)^{{\vec{\beta}}\cdot{\vec{\gamma}}}[g_c,[g_a,g_b\} \}= 0.&
\eea
The first equation gives the ${\mathbb Z}_2\times{\mathbb Z}_2$-graded (anti)commutators; the second equation
gives the ${\mathbb Z}_2\times{\mathbb Z}_2$-graded Jacobi identity. The generators $g_a, g_b, g_c$ respectively belong to the sectors ${\cal G}_{{\vec{\alpha}}},{\cal G}_{{\vec{\beta}}},{\cal G}_{{\vec{\gamma}}}$, where ${\vec{\alpha}}=(\alpha_1,\alpha_2)$ for $\alpha_{1,2}=0,1$
and ${\cal G}_{\vec\alpha}\equiv {\cal G}_{\alpha_1\alpha_2}$
(and similarly for ${\vec{\beta}}$, ${\vec{\gamma}}$). \par The scalar product ${\vec{\alpha}}\cdot{\vec{\beta}}$ is defined as
\bea
{\vec{\alpha}}\cdot{\vec{\beta}}&=& \alpha_1\beta_1+\alpha_2\beta_2.
\eea
For the (anti)commutator one has
$ [g_a,g_b\}\in {\cal G}_{{\vec{\alpha}}+{\vec{\beta}}}$, with the vector sum defined ${\textrm{mod}}~ 2$.\par
According to the definitions, the ${\mathbb Z}_2\times {\mathbb Z}_2$-graded  (anti)commutators $[A,B\}$ 
between two graded generators $A,B$ are read from the table
\bea\label{tableanticomm}
&\begin{array}{|c|c|c|c|c|}\hline 
 A\backslash B&00&10&01&11 \\  \hline
00&[\cdot,\cdot]&[\cdot,\cdot]&[\cdot,\cdot]&[\cdot,\cdot]\\ \hline
10&[\cdot,\cdot]&\{\cdot,\cdot\}&[\cdot,\cdot]&\{\cdot,\cdot\} \\  \hline
01&[\cdot,\cdot]&[\cdot,\cdot]&\{\cdot,\cdot\}&\{\cdot,\cdot\} \\  \hline
11&[\cdot,\cdot]&\{\cdot,\cdot\}&\{\cdot,\cdot\}&[\cdot,\cdot] \\ \hline
\end{array}.&
\eea
Let $V$ be a ${\mathbb Z}_2\times{\mathbb Z}_2$-graded vector space such that
\bea
{V}&=& {V}_{00}\oplus {V}_{10}\oplus{V}_{01}\oplus {V}_{11},
\eea
with $v,v'\in V$ of respective $ij$, $i'j'$ gradings. If $kl$ is the grading of the operator $M:V\rightarrow V$,
where $v'=Mv$, then we have, $mod ~2$,
\bea
i'=i+k, &\quad& j'=j+l.
\eea
In the paper we consider the  ``one-dimensional ${\mathbb Z}_2\times{\mathbb Z}_2$-graded supertranslation algebra" with generators
$H\in {\cal G}_{00}$, $Z\in {\cal G}_{11}$, $Q_{10}\in {\cal G}_{10}$, $Q_{01}\in {\cal G}_{01}$ and nonvanishing
(anti)commutators 
\bea\label{z2z2super}
\{Q_{10},Q_{10}\}=\{Q_{01},Q_{01}\}=2H,  &\quad& [Q_{10},Q_{01}] = -2Z.
\eea
The algebra
\bea\label{z2z2becdeb}
\relax \{Q_{10},Q_{10}\}=\{Q_{01},Q_{01}\}=2H,  &\quad& [Q_{10},Q_{01}] = [H,Q_{10}]=[H, Q_{10}]=0,
\eea 
with generators $H,Q_{10}, Q_{01}$ and vanishing commutator between $Q_{10},Q_{01}$ is referred to (following \cite{{BruDup},{BecDeb}}) as ``the Beckers-Debergh algebra". \par
In a $4\times 4$ matrix representation of the ${\mathbb Z}_2\times {\mathbb Z}_2$-graded superalgebra the 
nonvanishing entries (which can be either numbers or differential operators) of the ${\cal G}_{ij}$ graded sectors are  accommodated according to
\bea\label{gradedmatrices}
 {\cal G}_{00}= \left(\begin{array}{cccc}\ast&0&0&0\\0&\ast&0&0\\0&0&\ast&0\\0&0&0&\ast\end{array}\right), &\quad&
 {\cal G}_{11}= \left(\begin{array}{cccc}0&\ast&0&0\\\ast&0&0&0\\0&0&0&\ast\\0&0&\ast&0\end{array}\right),
\nonumber\\
{\cal G}_{10}= \left(\begin{array}{cccc}0&0&\ast&0\\0&0&0&\ast\\\ast&0&0&0\\0&\ast&0&0\end{array}\right),
 &\quad&
 {\cal G}_{01}= \left(\begin{array}{cccc}0&0&0&\ast\\ 0&0&\ast&0\\ 0&\ast&0&0\\ \ast&0&0&0\end{array}\right),
\eea
In the text the needed matrices are conveniently expressed as tensor products  of the $4$ real, $2\times 2$ split-quaternion matrices $I,X,Y,A$ given by  
{{
\bea\label{splitquat}
&I= \left(\begin{array}{cc}1&0\\0&1\end{array}\right),\quad X= \left(\begin{array}{cc}1&0\\0&-1\end{array}\right),\quad
Y= \left(\begin{array}{cc}0&1\\1&0\end{array}\right),\quad
A= \left(\begin{array}{cc}0&1\\-1&0\end{array}\right).&
\eea
}}

%
%
%

\end{document}